\documentclass[12pt]{article}
\usepackage{siunitx}
\usepackage{soul}
\usepackage{adjustbox}
\usepackage{threeparttable}
\usepackage{booktabs}

\usepackage{color}
\usepackage{graphics}
\usepackage{graphicx}
\usepackage{epsfig}
\usepackage{textcomp}
\usepackage[small]{caption} 
\usepackage{amsmath,amssymb}
\usepackage{tabularx}
\usepackage[table,usenames,dvipsnames]{xcolor}
\usepackage{hhline}
\usepackage{multicol}
\usepackage{multirow}
\newcolumntype{R}[1]{>{\raggedleft\let\newline\\\arraybackslash\hspace{0pt}}m{#1}} 
\usepackage{enumitem}
\usepackage[normalem]{ulem}
\usepackage{url} 
\usepackage[official]{eurosym}

\usepackage{xr}
\makeatletter
\newcommand*{\addFileDependency}[1]{
  \typeout{(#1)}
  \@addtofilelist{#1}
  \IfFileExists{#1}{}{\typeout{No file #1.}}
}
\makeatother
\newcommand*{\myexternaldocument}[1]{%
    \externaldocument{#1}%
    \addFileDependency{#1.tex}%
    \addFileDependency{#1.aux}%
}
\myexternaldocument{SI}

\usepackage{nameref}

\usepackage{pifont} 

\usepackage{times}

\title{Direct imaging of carbohydrate stereochemistry}

\author{Shuning Cai,$^{1\ast}$, Joakim S. Jestilä,$^{1\ast}$, Peter Liljeroth$^{1\dagger}$ and Adam S. Foster$^{1,4\dagger}$\\
\\
\normalsize{$^{1}$Department of Applied Physics, Aalto University, 00076 Aalto, Helsinki, Finland}\\
\normalsize{$^{4}$WPI Nano Life Science Institute (WPI-NanoLSI), Kanazawa University,}\\
\normalsize{Kakuma-machi, Kanazawa 920-1192, Japan}\\
\\
\normalsize{$^\ast$These authors contributed equally.}\\
\normalsize{$^\dagger$To whom correspondence should be addressed; }\\
\normalsize{E-mail: peter.liljeroth@aalto.fi; adam.foster@aalto.fi}
}

\begin{document}
\maketitle 
\baselineskip24pt

\begin{abstract}
Carbohydrates, essential biological building blocks, exhibit functional mechanisms tied to their intricate stereochemistry. Subtle stereochemical differences, such as those between the anomers maltose and cellobiose, lead to distinct properties due to their differing glycosidic bonds; the former is digestible by humans, while the latter is not. This underscores the importance of precise structural determination of individual carbohydrate molecules for deeper functional insights. However, their structural complexity and conformational flexibility, combined with the high spatial resolution needed, have hindered direct imaging of carbohydrate stereochemistry. Here, we employ non-contact atomic force microscopy integrated with a data-efficient, multi-fidelity structure search approach accelerated by machine learning integration to determine the precise 3D atomic coordinates of two carbohydrate anomers. We observe that glycosidic bond stereochemistry regulates on-surface chiral selection in carbohydrate self-assemblies. The reconstructed models, validated against experimental data, provide reliable atomic-scale structural evidence, uncovering the origin of on-surface chirality from carbohydrate anomerism. Our study confirms that nc-AFM is a reliable technique for real-space discrimination of carbohydrate stereochemistry at the single-molecule level, providing a pathway for bottom-up investigations into the structure-property relationships of carbohydrates in biological research and materials science.
\end{abstract}

\section{Introduction}
Carbohydrates, the most prevalent biomolecules, exhibit chirality—a fundamental property in chemistry, biology, physics, and material science—across multiple scales, ranging from sub- to supramolecular levels. Chirality conferral plays a critical role in the synthesis of chiral nanostructures with unusual optical and magnetic properties, as well as novel pharmaceutical compounds \cite{WOS:001106405600030, WOS:001281842200003}. However, the mechanism by which chirality transfers from individual molecules to self-assembled structures remains elusive \cite{morrow_transmission_2017, fittolani_bottom-up_2022,WOS:001230099000002}. Despite the fundamental importance of glycosidic bond stereochemistry in dictating biological processes such as enzyme recognition and cellular communication, its role in propagating chirality during self-assembly processes remains unexplained and lacks definitive structural evidence \cite{wang_synthesis_2015}. Carbohydrate function is highly dependent on subtle structural features, as the three-dimensional arrangement of atoms in carbohydrates is closely linked to their intrinsic stereoelectronic effects, including orbital interactions that govern both structure and function \cite{WOS:000285921600034, takezawa_enhanced_2020}. This emphasizes the need for precise structural determination at the submolecular level for unraveling the functional mechanisms of carbohydrates. 

The myriad of possible regio- and stereochemical combinations among more than one hundred known types of monosaccharides, combined with their extensive conformational flexibility, results in numerous possible isomers and intricate assembly rules for carbohydrate-based materials. The structural diversity not only offers carbohydrates high information density and a broad spectrum of tunable functionalities in both biological processes and synthetic materials \cite{djalali_towards_2024,delbianco_carbohydrates_2016}, but also complicates stereochemical control in carbohydrate synthesis and introduces substantial obstacles for structural characterization. This has contributed to the slower progress of the glycomics field compared to the rapid advancements in genomics and proteomics \cite{hofmann_identification_2015,hofmann_glycan_2017}. Established analytical techniques such as X-ray crystallography (XRD), cryogenic transmission electron microscopy (cryo-TEM), and nuclear magnetic resonance (NMR) face significant constraints when applied to carbohydrates due to poor crystallization, susceptibility to radiation damage (in TEM), and broad overlapping spectral signals resulting from the coexistence of multiple conformations \cite{kirschbaum_unravelling_2021,ogawa_recent_2022,delbianco_visualizing_2024}. Recent advances in scanning tunneling microscopy (STM) has achieved submolecular resolution real-space observation of individual carbohydrate molecules \cite{wu_imaging_2020,anggara_direct_2023, seibel_visualizing_2023}, opening new avenues for structural analysis of carbohydrates at the single-molecule level. Nevertheless, sub-nanometer resolution remains insufficient for fully resolving carbohydrate stereostructures. Providing the highest spatial resolution among real-space techniques, nc-AFM with a functionalized tip holds untapped potential for revealing stereochemical and conformational differences of individual carbohydrates \cite{gross_chemical_2009,bian_scanning_2021}.  Functionalizing the metal tip with a single, relatively inert atom or molecule (such as CO, Xe, or pentacene) enables very short tip-sample distances. As the tip approaches the surface, Pauli repulsion increasingly dominates, resulting in sharper image features. This allows for more precise sampling of molecular edges and for resolving individual chemical bonds, offering detailed insights into their structural configurations \cite{alldritt_automated_2020}.

As a surface analysis technique, nc-AFM is primarily applied to planar molecules and requires efficient structure search or reconstruction approaches when imaging non-planar molecules like carbohydrates, due to the non-intuitive images typically obtained when imaging 3D systems. While attempts have been made to employ machine learning approaches to infer 3D molecular structures directly from AFM images \cite{alldritt_automated_2020,priante_structure_2024}, these as yet remain too limited in accuracy when faced with the complexity of carbohydrate systems. To this end, we employ a data-efficient multi-fidelity global optimization protocol that merges active learning with density functional theory (DFT), which simultaneously provides training data for a machine learning interatomic potential (MLIP), which is then used to expand the number of candidate structures by accelerating their evaluation. As a result, here we demonstrate the real-space observation of carbohydrate stereochemistry in supramolecular assemblies with atomic-scale resolution by leveraging the aforementioned structure search protocol, validated by matching simulated and experimental constant-current STM and height-dependent nc-AFM images. The precise atomic positions of our validated models enable correlating the distinct on-surface chiralities of the self-assemblies with the stereoelectronic properties of the individual molecular building blocks, providing a rare example of homochiral self-assemblies induced by carbohydrate anomerism. Our work provides a bottom-up approach to studying structure-property relationships in carbohydrate molecules, opening an avenue for the development of carbohydrate-based supramolecules, ultimately shedding light on the design of complex molecular architectures.

\section{Results and discussion}
\subsection{The Role of Anomerism in Carbohydrate Self-Assembly on Au(111) Surfaces}

\begin{figure*}[t!]
    \centering
    \includegraphics[width=\textwidth]{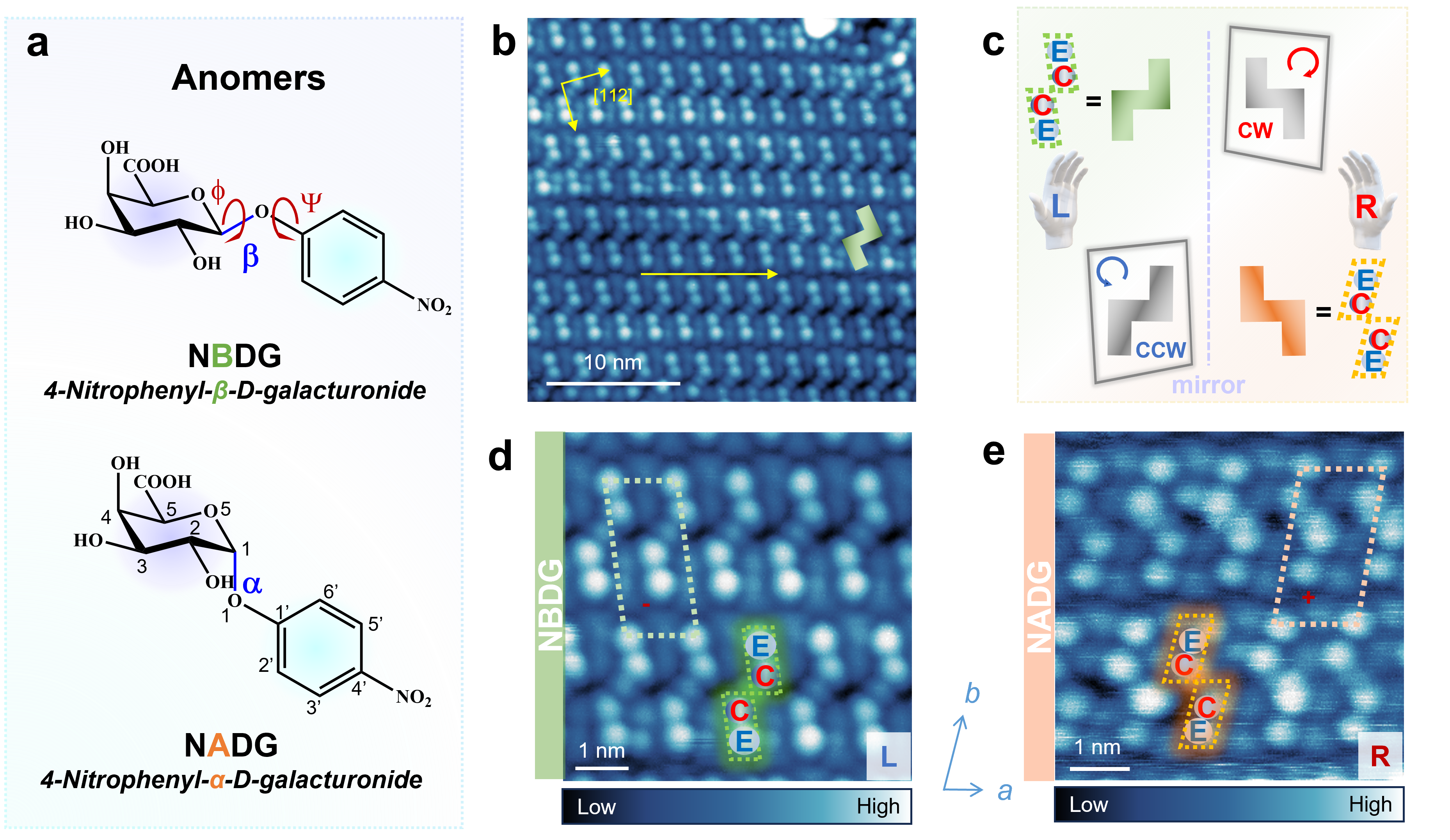}
    \caption{\textbf{Influence of anomerism on chiral selection in carbohydrate self-assembly on Au(111).} (\textbf{a}) Chemical structure of 4-nitrophenyl-$\alpha$-D-galacturonide \textbf{NBDG} (top) and 4-nitrophenyl-$\beta$-D-galacturonide \textbf{NADG} (bottom), which differ in the orientation of the glycosidic bond (C$_1$--O$_1$) highlighted in navy, marked with $\alpha$ and $\beta$ symbols, respectively. (\textbf{b}) High-resolution STM topography probed by a CO-functionalized tip of the \textbf{NBDG} self-assembly. Set point: 200 mV, 5 pA. (\textbf{c}) Schematic representation of the chiral relationship of the \textbf{NBDG} and the \textbf{NADG} lattice basic unit. \textbf{d}, \textbf{e}, Zoomed-in high-resolution STM topography of \textbf{NBDG} (\textbf{d}) and \textbf{NADG} (\textbf{e}) self-assemblies. Set point: 200 mV, 5 pA; 247 mV, 5.76 pA, respectively.}
    \label{STMLargescale}
\end{figure*}

To investigate the impact of carbohydrate stereochemistry on 2D crystallization, we studied two carbohydrate molecules: 4-nitrophenyl-$\beta$-D-galacturonide (\textbf{NBDG}) and 4-nitrophenyl-$\alpha$-D-galacturonide (\textbf{NADG}), as depicted in the chemical structures in Figure \ref{STMLargescale}a. The two molecules were deposited sequentially onto the Au(111) surface. Both \textbf{NBDG} and \textbf{NADG} possess the same monosaccharide backbone and nitrophenyl substituent marked in \textcolor{violet}{lilac} and \textcolor{cyan}{cyan}, respectively. The key difference between the two molecules is the connectivity and orientation of the glycosidic bond (C$_{1}$--O$_{1}$, highlighted in \textcolor{blue}{navy}), which is part of the linkage connecting the two subunits,  resulting in their classification as anomers. In the $\beta$-anomer, the exocyclic oxygen (O$_{1}$) is bonded to the anomeric carbon (C$_{1}$) in the equatorial position, whereas in the $\alpha$-anomer, the attachment is in the axial position. It is important to note that the C$_{1}$--O$_{1}$ and O$_{1}$--C$_{1}$’ bonds, along with the --OH and --COOH groups in the monosaccharide unit, are rotatable, which enables multiple conformations. 

\begin{figure}[htbp]%
    \centering
    \includegraphics[width=0.8\textwidth]{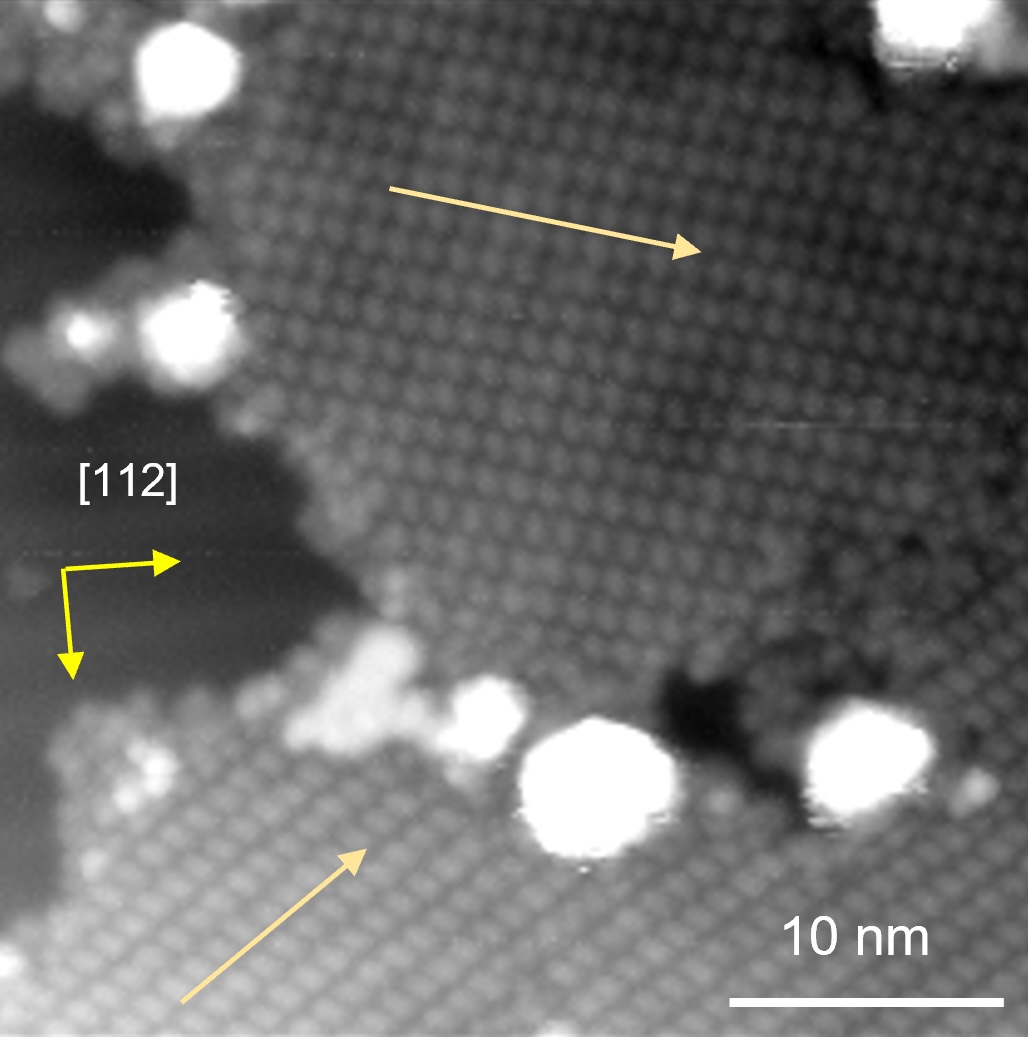}
    \caption{\textbf{The large-scale STM of \textbf{NBDG}.} The \textit{a}-axis vector of the unit cell, indicated by the yellow arrow, is oriented at $-15.5 \pm 1$\textdegree{} and $+35.5 \pm 1$\textdegree{} relative to the [$11\overline{2}$] direction of the substrate, represented by the white arrow, for the upper and lower domains, respectively.}
    \label{fig:beta_largescale}
\end{figure}

Following the ESD of \textbf{NBDG} from ambient conditions into UHV at room temperature, ordered homochiral superstructures emerge on the Au(111) surface (Figure~\ref{STMLargescale}b and Figure~\ref{fig:beta_largescale}). We notice that the growth orientation of the \textbf{NBDG} self-assemblies shows a weak correlation with the lattice directions of the Au(111) surface. This is suggested by the angles between the \textit{a}-axis of the unit cell and the [$11\overline 2$] direction of the substrate across two different domains, as illustrated in yellow arrow, which are $-15.5 \pm 1$\textdegree{} and $+35.5 \pm 1$\textdegree{} (see Figure~\ref{fig:beta_largescale}). In the high-resolution STM probed by a CO tip (Figure \ref{STMLargescale}b), we observe that the basic unit of the lattice consists of four bright protrusions, depicted in a green Z-shape pattern. In the enlarged high-resolution STM image (Figure \ref{STMLargescale}d), the two central protrusions, labeled as \textcolor{red}{\textbf{C}}, are identical to each other but differ from the edge protrusions, labeled as \textcolor{blue}{\textbf{E}}, the latter two also being identical to each other.  

\begin{figure}[htbp]%
    \centering
    \includegraphics[width=\textwidth]{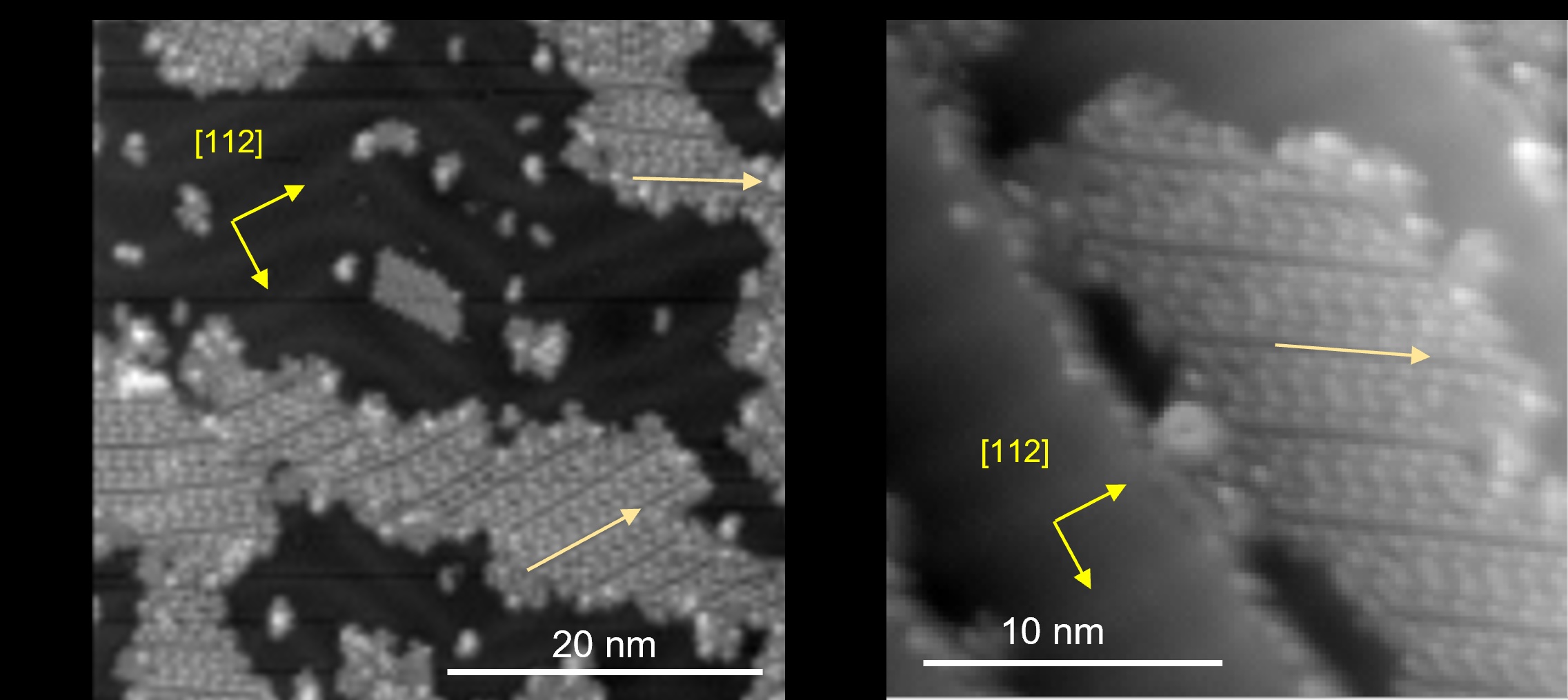}
    \caption{\textbf{The large-scale STM of \textbf{NADG} taken from different area.} \textbf{a}, The \textit{a}-axis vector of the unit cell, indicated by the yellow arrow, is oriented at $+30.0 \pm 1$\textdegree{} and $0.0 \pm 1$\textdegree{} relative to the [$11\overline{2}$] direction of the substrate, represented by the white arrow, for the upper and lower domains, respectively. \textbf{b}, The \textit{a}-axis vector of the unit cell is oriented at $+30.0 \pm 1$\textdegree{} relative to the [$11\overline{2}$] direction of the substrate.}
    \label{fig:alpha_largescale}
\end{figure}

In similar fashion, \textbf{NADG} forms regular self-assemblies on Au(111) upon deposition at room temperature, as shown in Figure~\ref{STMLargescale}e and Figure~\ref{fig:alpha_largescale}. In contrast, the growth orientation of the \textbf{NADG} self-assemblies appears to correlate with the lattice directions. Different domains exhibit the \textit{a}-axis vector aligning either along the [$11\overline 2$] direction or forming a 30\textdegree{} angle with it (see Figure~\ref{fig:alpha_largescale} for more details). The basic unit of the \textbf{NADG} lattice, depicted within the orange frame, also comprises two pairs of bright protrusions. Like the \textbf{NBDG} basic unit, the central and edge protrusions, denoted as \textcolor{red}{\textbf{C}} and \textcolor{blue}{\textbf{E}} respectively, are mutually identical. Unlike the negative angle observed in the \textbf{NBDG} unit cell, the angle between the \emph{a}-axis relative to the \emph{b}-axis is positive in the \textbf{NADG} lattice (see Figure \ref{fig:alpha_largescale} for further details). The Z(S)-shaped basic \textbf{NADG}/\textbf{NBDG} unit exhibits  \textit{C$_{2}$} point group symmetry, while non-superimposability with the mirror image indicates chirality, as illustrated in Figure \ref{STMLargescale}c.  Moreover, the basic unit of \textbf{NADG} exhibits right-handedness (R), whereas that of \textbf{NBDG} displays left-handedness (L). The propagation of the chiral \textbf{NADG}/\textbf{NBDG} basic unit throughout the lattice imparts consistent chirality to the entire superstructure. Given that self-assemblies of the opposite chirality are not observed, this phenomenon demonstrates spontaneous chiral selection on an achiral surface for the two carbohydrate anomers.

\subsection{Reconstruction of 3D atomic-resolution structural models}

\begin{figure*}[t!]
    \centering
    \includegraphics[width=\textwidth]{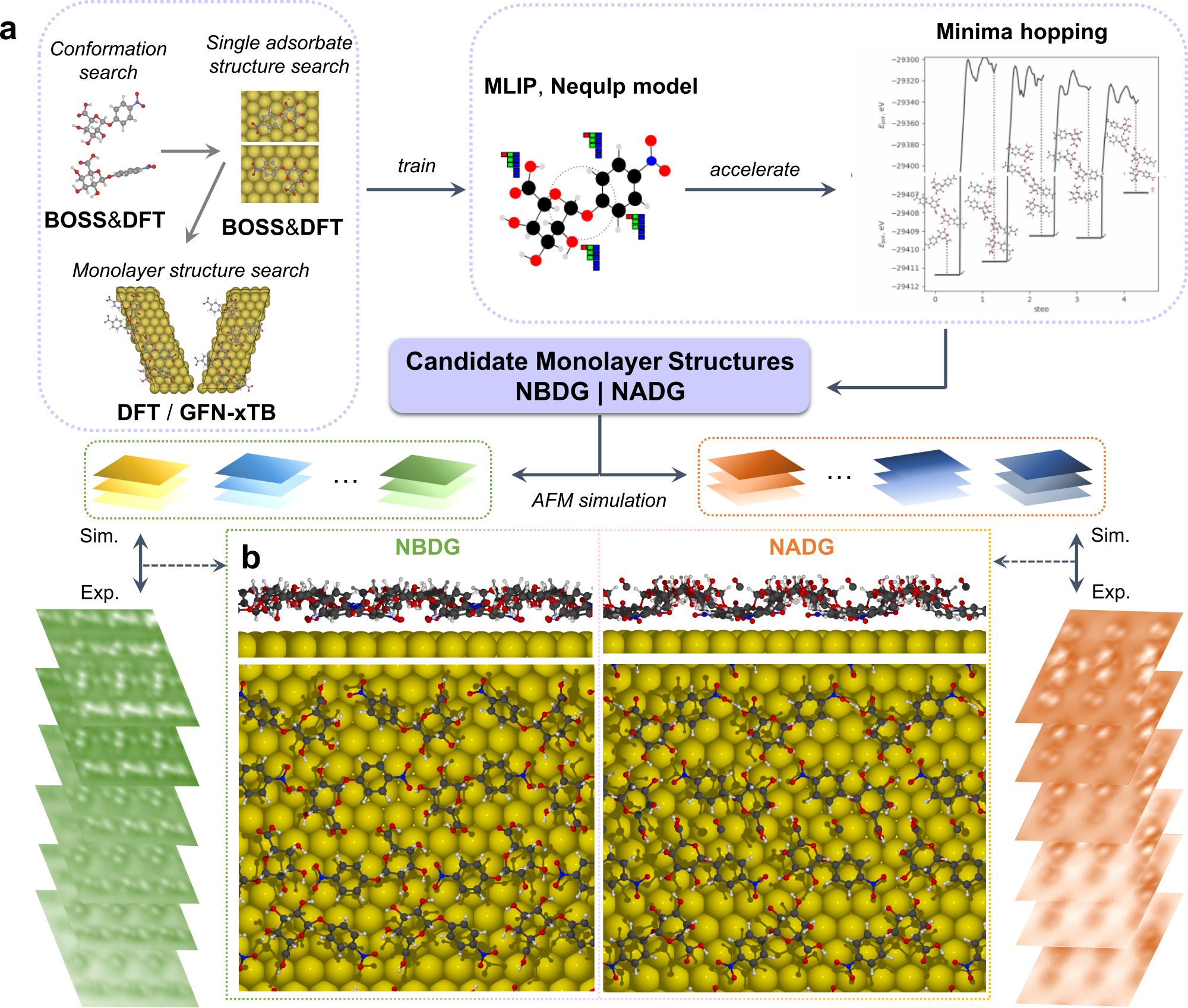}
    \caption{\textbf{Workflow of 3D atomic-resolution structural model reconstruction. } \textbf{a},  Monolayer adsorption configuration structure search employing a multi-fidelity modeling approach (left to right): the search begins by identifying the most stable conformers and adsorption configuration for single-molecule \textbf{NBDG}/\textbf{NADG} adsorbates, which is achieved using a surrogate model of the DFT potential energy surface constructed via BOSS to accelerate the identification of the global minima; Initial monolayer structures are manually constructed and relaxed using high-fidelity DFT, supplemented by high-temperature molecular dynamics snapshots from the GFN-xTB method. The aforementioned acquired data are subsequently used to train a MLIP, NequIP. The trained potential accelerates the exploration of possible structures with minima hopping, providing a set of candidate monolayer structures.  \textbf{b}, The final monolayer structures of \textbf{NBDG} and \textbf{NADG} are obtained by manually adjusting the global minimum structures from the minima hopping procedure until the simulated AFM images resemble the experimental counterparts, the resulting structures being the most consistent with the experimental results.}
    \label{methods_structure_determination}
\end{figure*}

To resolve the underlying atomic-resolution structures of the \textbf{NBDG} and \textbf{NADG} self-assemblies, we conducted height-dependent AFM measurements with a CO-functionalized tip (Figure~\ref{methods_structure_determination}). During the initial structural screening, we observed that even small changes in the 3D monolayer structures of \textbf{NBDG}/\textbf{NADG} are distinctly reflected in the simulated AFM images, demonstrating the potential of AFM to guide detailed structural investigations. 

Achieving agreement between a structural model and experimental data requires extensive exploration of possible structures, in particular when considering a full monolayer consisting of flexible 3D molecules. To address the labor-intensive construction of monolayer structures and the computational cost of subsequent DFT computations, we employ the minima hopping algorithm \cite{goedecker_minima_2004}, integrated with a state-of-the-art MLIP, the highly data-efficient Neural Equivariant Interatomic Potential (NequIP) \cite{batzner_e3-equivariant_2022}. A simplified schematic of the workflow of the 3D atomic-resolution structural model reconstruction is presented in Figure \ref{methods_structure_determination}. Both the conformational search and the adsorption configuration search for an individual \textbf{NBDG}/\textbf{NADG} molecule are performed on a surrogate model of the DFT potential energy surface (PES), constructed with the Bayesian Optimization Structure Search (BOSS) package \cite{todorovic_bayesian_2019}. Initial monolayer structures are manually built using the most stable single-molecule adsorption configurations, guided by the lattice parameters derived from experimental STM images (Figures \ref{fig:alpha_lines}, \ref{fig:beta_lines}, and \ref{fig:isolated_molecules_on_surface}), and then relaxed using DFT. The aforementioned data is supplemented with high-temperature molecular dynamics snapshots from the semiempirical tight-binding DFT method GFN-xTB \cite{grimme_general_2017}, which are then used to train the NequIP model. While BOSS is limited to less than 20 degrees of freedom, making it insufficient for full monolayer structure establishment, minima hopping using the trained NequIP model overcomes this limitation, accelerating the sampling process without compromising computational accuracy. This multi-fidelity modeling approach was used to obtain 850 candidate monolayer structures for the two anomers. The simulated AFM stacks of the global minimum structures are overall close to the experimental results and show better agreement than candidates corresponding to higher energies. The final reconstructed 3D monolayer structures of \textbf{NBDG} and \textbf{NADG} (Figure \ref{methods_structure_determination}b) are obtained by refining the atomic positions of the global minimum structures, followed by final constrained DFT relaxations. 

\begin{figure*}[htbp]
    \centering
    \includegraphics[width=\textwidth]{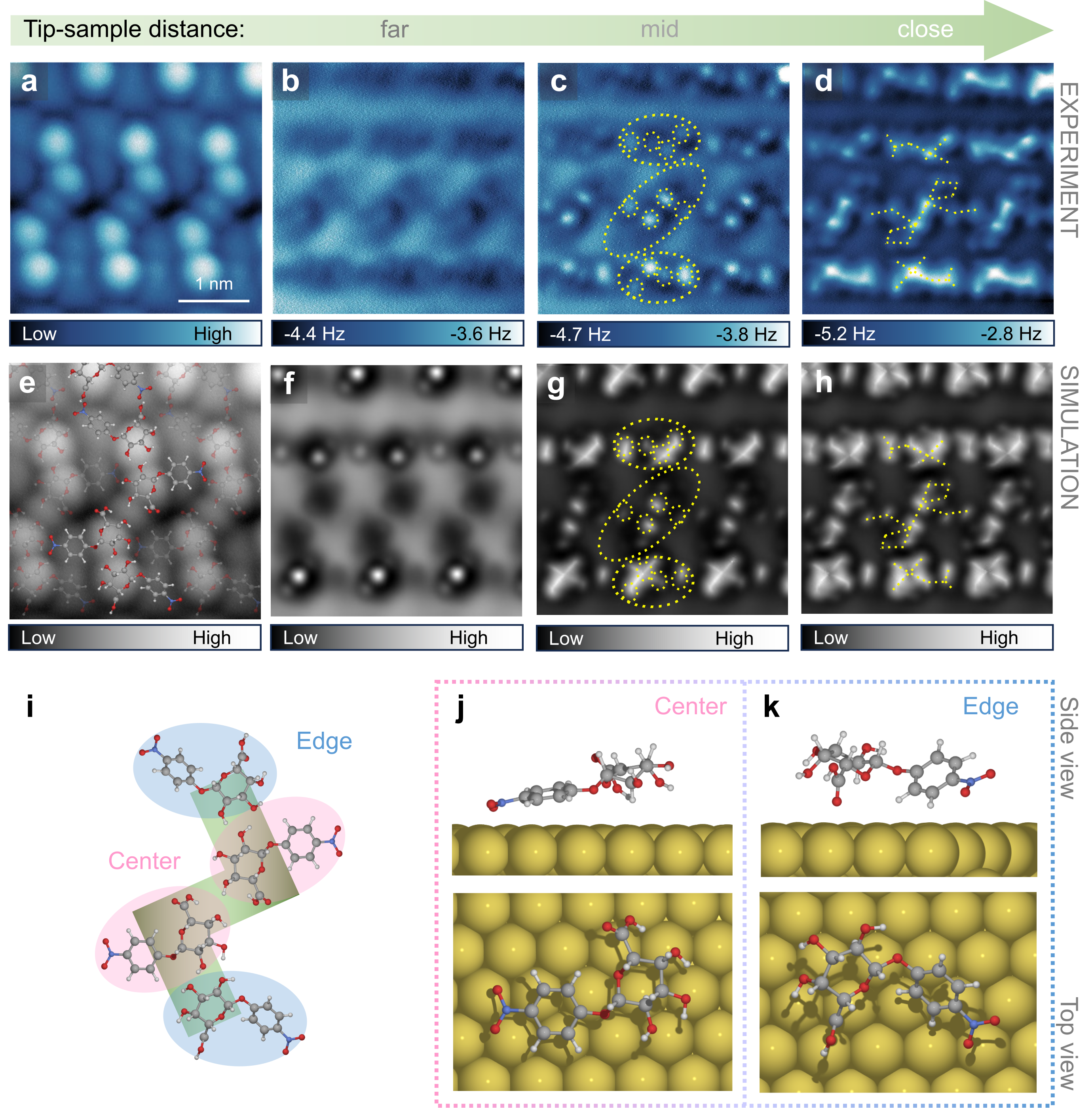}]
    \caption{\textbf{STM and height-dependent AFM images of the \textbf{NBDG} self-assembly.} \textbf{a}, \textbf{e}, High-resolution CO-tip STM images covering the same area as in the height-dependent AFM stack \textbf{(a)}, the second image showing the superimposed 3D reconstructed molecular structure \textbf{(e)}. Set point: 200mV, 5pA. \textbf{b}-\textbf{d}, Constant-height AFM images at different tip-sample distances, far: \textminus80pm \textbf{(b)}; mid: \textminus50pm \textbf{(c)}; and close: \textminus20pm \textbf{(d)}. The tip height values are relative to the STM set point (200mV, 5pA). \textbf{f}-\textbf{h}, The simulated AFM images approximately corresponding to the experimental far \textbf{(f)}, mid \textbf{(g)} and close tip-sample distances \textbf{(h)}. \textbf{i}-\textbf{k}, The absolute configuration of the \textbf{NBDG} basic unit, displaying two distinct conformations: the central (pink) and the edge conformations (blue) \textbf{(i)}, The top view (upper panels) and side view (lower panels) of the individual central \textbf{(j)} and edge \textbf{(k)} conformations adsorbed on Au(111). Yellow dashed lines are added to guide eye towards matching features in the experimental and simulated images.}
    \label{beta_AFM_STM}
\end{figure*}

\begin{figure*}[htbp]
    \centering
    \includegraphics[width=\textwidth]{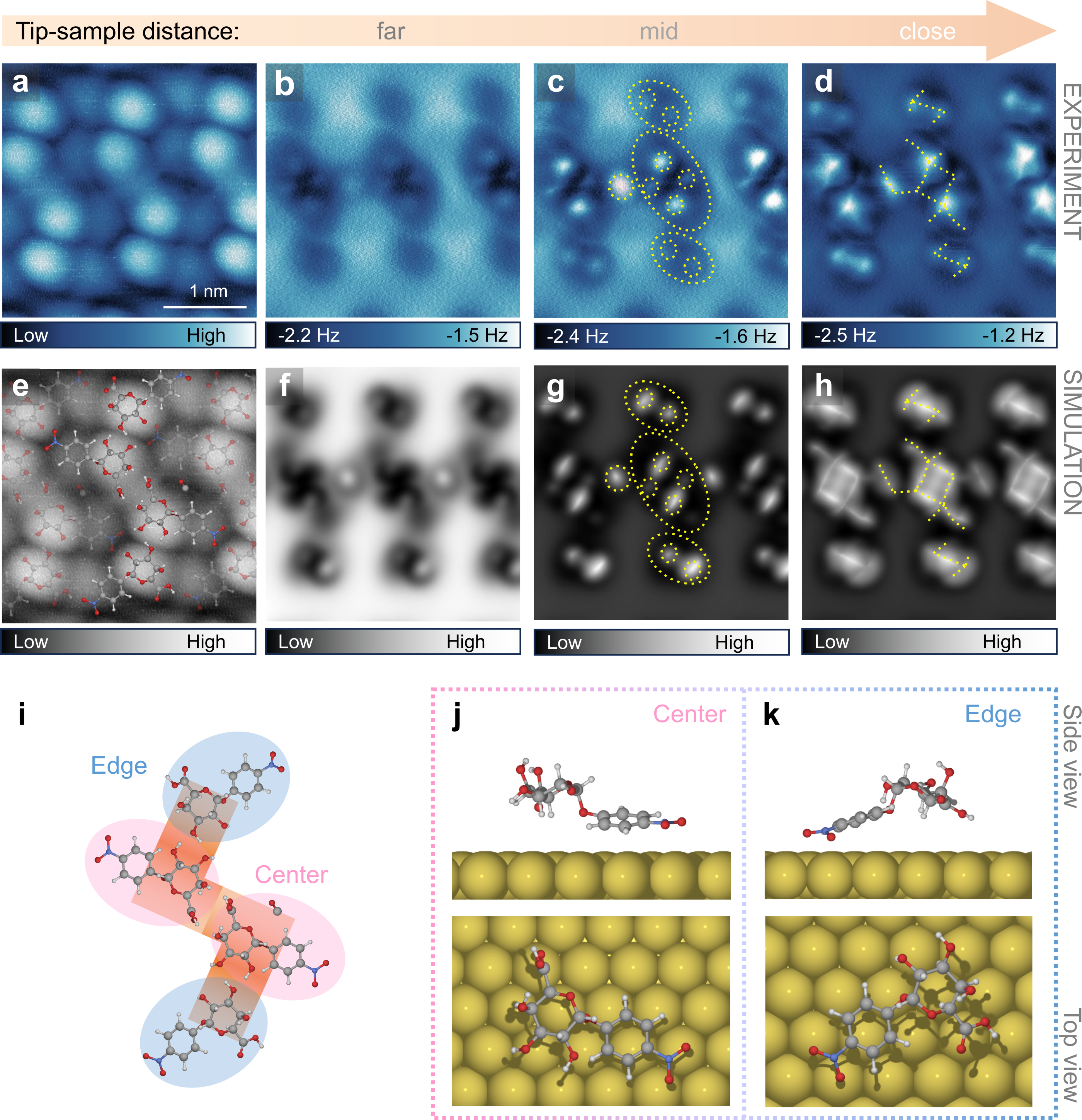}
    \caption{\textbf{STM and height-dependent AFM images of the \textbf{NADG} self-assembly.} \textbf{a}, \textbf{e}, High-resolution CO-tip STM images covering the same area as in the height-dependent AFM stack \textbf{(a)}, the second image showing the superimposed 3D reconstructed molecular structure \textbf{(e)}. Set point: 247mV, 5.76pA. \textbf{b}-\textbf{d}, Constant-height AFM images at different tip-sample distances, far: \textminus90pm \textbf{(b)}; mid: \textminus70pm \textbf{(c)}; and close: \textminus50pm \textbf{(d)}. The tip height values are relative to the STM set point (200mV, 5pA). \textbf{f}-\textbf{h}, The simulated AFM images approximately corresponding to the experimental far \textbf{(f)}, mid \textbf{(g)} and close tip-sample distances \textbf{(h)}. \textbf{i}-\textbf{k}, The absolute configuration of the \textbf{NADG} basic unit, displaying two distinct conformations: the central (pink) and the edge conformations (blue) \textbf{(i)}, The top view (upper panels) and side view (lower panels) of the individual central \textbf{(j)} and edge \textbf{(k)} conformations adsorbed on Au(111). Yellow dashed lines are added to guide eye towards matching features in the experimental and simulated images.}
    \label{alpha_AFM_STM}
\end{figure*}

The simulated AFM stacks at different tip-sample distances based on the reconstructed 3D models of the \textbf{NBDG} and \textbf{NADG} molecules (Figures \ref{beta_AFM_STM}f -- h and \ref{alpha_AFM_STM}f -- h), closely align with the corresponding experimental results, supporting the validity of the model structures. Overlaying the \textbf{NBDG} and \textbf{NADG} structures onto their STM images (Figure \ref{beta_AFM_STM}a, e and Figure \ref{alpha_AFM_STM}a, e), captured in the same area as the AFM stacks, reveals that both the left-handed \textbf{NBDG} basic unit and the right-handed \textbf{NBDG} basic unit comprises four molecules with two distinct conformations: denoted as the central and edge conformations, which are highlighted in \textcolor{pink}{pink} and \textcolor{blue}{blue}, respectively (Figure \ref{beta_AFM_STM}i and Figure \ref{alpha_AFM_STM}i). The bright protrusions observed in the STM images are primarily attributed to the pyranose rings of the \textbf{NBDG} and \textbf{NADG} molecules. In both the central and edge conformations for the \textbf{NBDG} basic unit, the --O$_{3}$H, --O$_{4}$H, and --COOH groups in the pyranose ring are oriented toward the Au(111) surface, as clearly seen in the individual adsorption configurations from the top and side views (Figure \ref{beta_AFM_STM}j and k).  In contrast, for the \textbf{NADG} basic unit, these same groups are oriented away from the Au(111) surface, while the glycosidic bond (C${_1}$--O${_1}$) remains oriented toward the surface, similar to the \textbf{NBDG} conformations, leading to a more parallel alignment between the nitrophenyl group and the surface (Figure \ref{alpha_AFM_STM}j and k). Additionally, DFT calculations indicate that the rotational barriers of the --OH groups in the central units of the \textbf{NADG} assembly are relatively low, around 10 kJ/mol at their minimum (see SI, Figure \ref{fig:alpha_OH_rotational_energy_analysis}), compared to the more rigid hydrogen atoms in the pyranose ring. This agrees with our experimental observations, suggesting that the --OH groups are more susceptible to movement due to tip-sample interactions. At closer tip-sample distances, the simulated AFM images for \textbf{NBDG} show better agreement with experimental data, as the contrast features predominantly arise from the rigid hydrogen atoms (Figure \ref{beta_AFM_STM}j and k, side view). In contrast, for \textbf{NADG}, the --OH groups significantly contribute to the contrast features (Figure \ref{alpha_AFM_STM}j and k, side view), leading to deviations between the simulated and experimental images at close tip-sample distances. These deviations arise from changes in the atomic positions of the more flexible --OH groups caused by the tip-sample interactions, which are not accounted for in the AFM simulations (more details in Figure \ref{fig:alpha_OH_rotational_energy_analysis}).  

Notably, despite the presence of multiple on-surface configurations, the molecules exhibit the same on-surface chirality in the self-assemblies of both \textbf{NBDG} and \textbf{NADG}, as indicated by the opposite orientation of the --O$_{3}$H, --O$_{4}$H, and --COOH groups in the pyranose rings relative to the surface, as well as the similar orientation of the exocyclic oxygen (O$_1$) toward the substrate. These groups tend to follow a similar hydrogen bonding pattern, resulting in the molecules being arranged in an alternating orientation sequence within the basic unit on the surface. This reveals that both \textbf{NBDG} and \textbf{NADG} molecules in their respective hydrogen-bonded self-assemblies exhibit chiral-selective adsorption, leading to left-handedness and right-handedness in the basic unit and overall superstructures, respectively.

\section{Stereoelectronic effects of individual carbohydrate \\ molecules on the assembled structures}

\begin{figure*}[htbp]
    \centering
    \includegraphics[width=\textwidth]{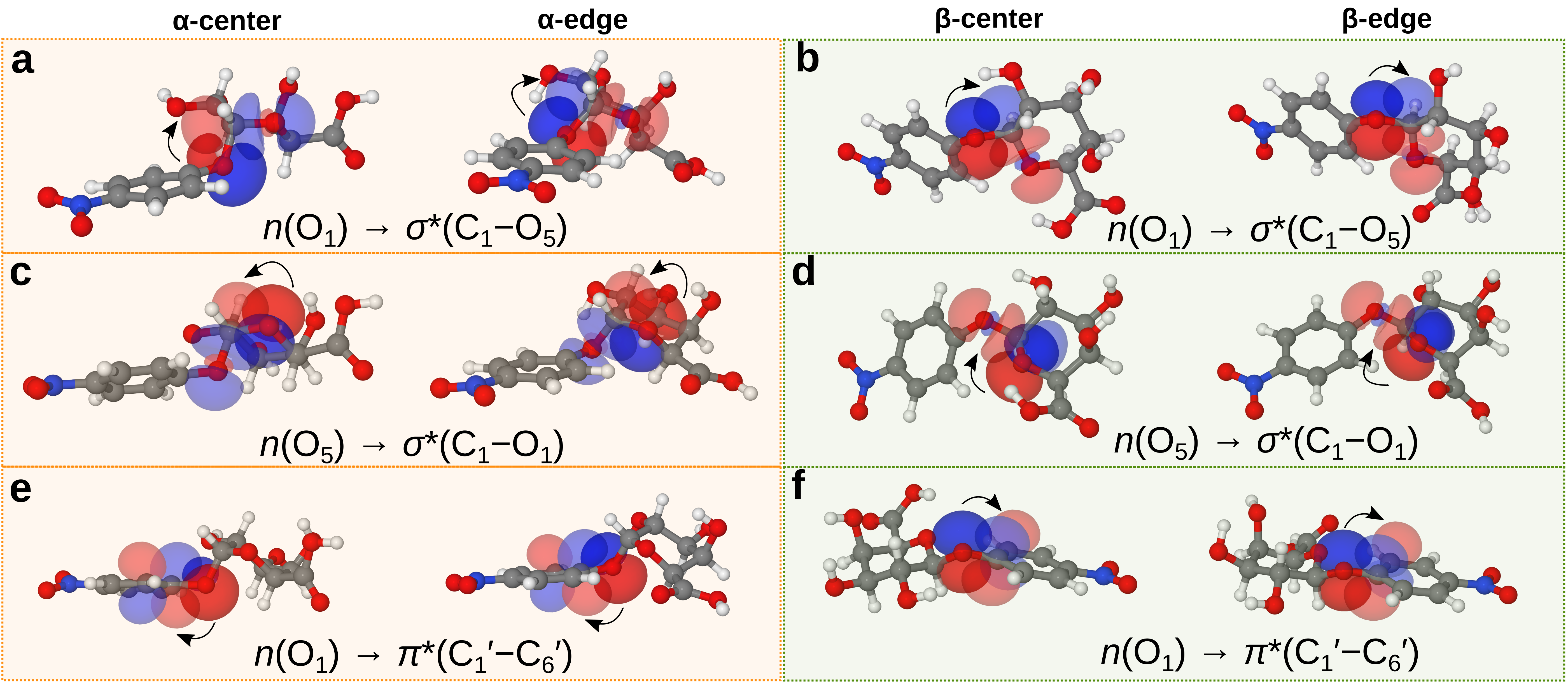}
    \caption{\textbf{Selected donor-acceptor orbitals relevant to the stereoelectronics of individual molecules in the \textbf{NADG}/\textbf{NBDG} assemblies}. Donor-acceptor orbital pairs are superimposed onto the same molecule, with the empty acceptor orbitals being slightly more translucent than the filled donor orbitals, while the electron flow direction is indicated with curved arrows. \textbf{NADG} structures are displayed with orange background, \textbf{NBDG} with green background, both showing the central (left) and edge conformations (right), respectively. \textbf{a},\textbf{b}, Donor-acceptor interactions related to the exo-anomeric effect for \textbf{NADG} (\textbf{a}) and \textbf{NBDG} (\textbf{b}). \textbf{c},\textbf{d}, Donor-acceptor interactions related to the endo-anomeric effect for \textbf{NADG} (\textbf{c}) and \textbf{NBDG} (\textbf{d}). \textbf{e},\textbf{f}, Donor-acceptor interactions related to the donation of electron density to the nitrophenyl substituent for \textbf{NADG} (\textbf{e}) and \textbf{NBDG} (\textbf{f}).}
    \label{donor_acceptor_NBOs}
\end{figure*}

\begin{table}[ht]
    \centering
    \small
        \begin{tabular}{ccccc}
        \toprule
         & $\alpha$-center  & $\alpha$-edge &  $\beta$-center* & $\beta$-edge* \\
         \midrule
         \textit{d}(C$_{1}$--O$_{5}$) & 1.41 & 1.42 & 1.44 & 1.42 \\
         \textit{d}(C$_{1}$--O$_{1}$) & 1.47 & 1.44 & 1.41 & 1.42 \\
         \hline
        $\phi$ ($\angle$O$_5$--C$_1$--O$_1$--C$_1$’) & 134.5 & 90.0 & 279.1 & 294.5 \\
        $\psi$ ($\angle$C$_1$--O$_1$--C$_1$'--C$_6$’) & 93.7 & 125.4 & 225.1 & 217.8 \\
        \hline
        \textbf{exo:} \textit{n}$_{(O1)}$ $\rightarrow$ $\sigma$*$_{(C1-O5)}$ & 24.1 & 49.2 & 57.9 & 47.3 \\
         \textbf{endo:} \textit{n}$_{(O5)}$ $\rightarrow$ $\sigma$*$_{(C1-O1)}$ & 59.8 & 65.6 & 15.4 & 16.4 \\
        \hline
        $\Sigma$ $\rho$(r) $\rightarrow$ $\sigma$*$_{(C1-O5)}$ & 58.9 & 82.0 & 93.6 & 81.0\\
        $\Sigma$ $\rho$(r) $\rightarrow$ $\sigma$*$_{(C1-O1)}$ & 118.7 & 111.9 & 67.5 & 75.3\\
         \textit{n}$_{(O1)}$ $\rightarrow$ $\sigma$*$_{(C1'-C2')}$ & 27.9 & 14.8 & 10.0 & 10.1 \\
         \textit{n}$_{(O1)}$ $\rightarrow$ $\sigma$*$_{(C1'-C6')}$ & 28.9 & 31.9 & 32.3 & 33.8 \\
         \textit{n}$_{(O1)}$ $\rightarrow$ $\pi$*$_{(C1'-C6')}$ & 32.8 & 56.2 & 74.3 & 87.4 \\
         \bottomrule
    \end{tabular}
    \caption{Selected computational bond distances in Å, dihedral angles $\phi$, and donor-acceptor interaction energies from second order perturbation theory in the NBO basis for the \textbf{NADG}/\textbf{NBDG} adsorbates in kJ/mol. *\textbf{NBDG} parameters from fully relaxed structure}
    \label{tab:table_1}
\end{table}

Having obtained precise structural models of the monolayers, we are in a position to discern their intrinsic stereoelectronic effects. This is achieved through a natural bond orbital (NBO) analysis, shedding light on the intra- and intermolecular interactions underlying the observed chiral selection. The analysis (Figure \ref{donor_acceptor_NBOs}, Table \ref{tab:table_1}) reveals vital aspects of orbital interactions related to stereoelectronic effects, such as those related to the endo- and exo-anomeric effects, which play an important role in determining the molecular structure of carbohydrates. These two effects correspond to \textit{n}${(O_5)}$ $\rightarrow$ $\sigma$*${(C_1-O_1)}$ and \textit{n}${(O_1)}$ $\rightarrow$ $\sigma$*${(C_1-O_5)}$ hyperconjugative interactions, respectively \cite{cocinero_sensing_2011}. Generally, a donor-acceptor interaction with a higher energy has a larger effect on the overall molecular structure than interactions with lower energies. In \textbf{NADG}, the endo-anomeric effect dominates in both the $\alpha$-center and $\alpha$-edge conformations (Figure \ref{donor_acceptor_NBOs}c). However, the relative contribution of the exo-anomeric effect increases substantially in the $\alpha$-edge, as indicated by the reduced energy difference between the two hyperconjugative interactions (\textit{$\Delta$E} = 35.7 kJ/mol $\rightarrow$ 16.3 kJ/mol, for $\alpha$-center and $\alpha$-edge, respectively). A dihedral angle change of 44.5° (O$_5$--C$_1$--O$_1$--C$_1$’) in going from $\alpha$-center to $\alpha$-edge further supports improved orbital alignment between \textit{n}${(O_1)}$ and $\sigma$*${(C_1-O_5)}$ (Figure \ref{donor_acceptor_NBOs}a). Additionally, in the $\alpha$-center, the C$_{1}$--O$_{5}$ bond is 0.06 Å shorter than C$_{1}$--O$_{1}$, whereas this same bond length difference reduces to 0.02 Å in the $\alpha$-edge, in line with the more balanced contributions of endo- and exo-anomeric effects.

In \textbf{NBDG}, the exo-anomeric effect dominates for both the $\beta$-center and $\beta$-edge conformations, while the endo-anomeric effect is deactivated in the equatorial conformation due to the misalignment between the \textit{n}${(O_5)}$ lone pair and the $\sigma$*${(C_1-O_1)}$ orbital, as indicated by the $\phi$ dihedral angles (Figure \ref{donor_acceptor_NBOs}b and d). The interaction energies reflect this deactivation, with the exo-anomeric effect providing the most favorable interactions at 57.9 kJ/mol and 47.3 kJ/mol in the $\beta$-center and $\beta$-edge, respectively, compared to the endo-effect at 15.4 kJ/mol (\textit{$\Delta$E} = 42.5 kJ/mol) and 16.4 kJ/mol (\textit{$\Delta$E} = 30.9 kJ/mol). Notably, the C$_{1}$--O$_{5}$ and C$_{1}$--O$_{1}$ bond length difference is less pronounced in \textbf{NBDG} compared to \textbf{NADG}, particularly for the $\beta$-edge conformation, where both bonds measure 1.42 Å. The endo and exo-anomeric effects alone cannot explain this, as the difference in interaction energies here is comparable to that of $\alpha$-edge, which \textit{does} display different C$_{1}$--O$_{5}$ and C$_{1}$--O$_{1}$ bond lengths. This discrepancy can be understood by including the sum of interaction energies between all donors and the corresponding antibonding orbitals in the analysis. These bond distances being equal, we should expect a minor difference in the total interaction energies for the two antibonding orbitals, $\Sigma$ $\rho$(r) $\rightarrow$ $\sigma$*${(C_1-O_5)}$ and $\Sigma$ $\rho$(r) $\rightarrow$ $\sigma$*${(C_1-O_1)}$. The total donation to $\sigma$*${(C_1-O_5)}$ is larger in the $\beta$-center (\textit{$\Delta$E} = 12.6 kJ/mol), while the donation to $\sigma$*${(C_1-O_1)}$ is smaller (\textit{$\Delta$E} = \textminus7.8 kJ/mol) than in the $\beta$-edge, increasing the interaction energy difference between the two antibonding orbitals (\textit{$\Delta$E} = 26.1 kJ/mol) compared to the $\beta$-edge (\textit{$\Delta$E} = 5.7 kJ/mol). In contrast, the interaction energy difference is even more pronounced in \textbf{NADG}, particularly in the $\alpha$-center, reaching up to 59.8 kJ/mol, aligning with the greater variation in C$_{1}$--O$_{5}$ and C$_{1}$--O$_{1}$ bond lengths. Furthermore, the aromatic nitrophenyl ring, which can interact with the lone pair orbitals on the glycosidic oxygen O$_1$ through the $\pi$* and $\sigma$* orbitals, can influence the total interaction energies and the charge distribution \cite{singh_direct_2016}. The total donation from \textit{n}${(O_1)}$ to the phenyl ring is more pronounced in \textbf{NBDG}, with the primary contribution arising from the interaction with the $\pi$${(C_1'-C_6')}$ orbital, especially in the $\beta$-edge conformation from \textit{n}${(O_1)}$ $\rightarrow$ $\pi$*${(C_1'-C_6')}$ at 87.4 kJ/mol (Figure \ref{donor_acceptor_NBOs}f). The same interaction is also significant in the $\beta$-center at 74.3 kJ/mol, compared to the lower values in the $\alpha$-center and $\alpha$-edge, at 32.8 kJ/mol and 56.2 kJ/mol, respectively (Figure \ref{donor_acceptor_NBOs}e). This pattern aligns with the $\Sigma$ $\rho$(r) $\rightarrow$ $\sigma$${(C_1-O_5)}$ and $\Sigma$ $\rho$(r) $\rightarrow$ $\sigma$*${(C_1-O_1)}$ interaction energies and corresponding bond length variations. The interactions between the \textit{n}${(O_1)}$ lone pair with the nitrophenyl group compete with those between the \textit{n}${(O_1)}$ lone pair and the C$_1$--O$_5$, counteracting the latter by providing an additional pathway for electron density transfer. This results in a more balanced charge distribution and a reduced difference in bond lengths. More details on the individual donor-acceptor orbitals in the SI (Figure \ref{fig:orbital_interactions}).

\begin{figure}[htbp]
    \centering
    \includegraphics[width=\textwidth]{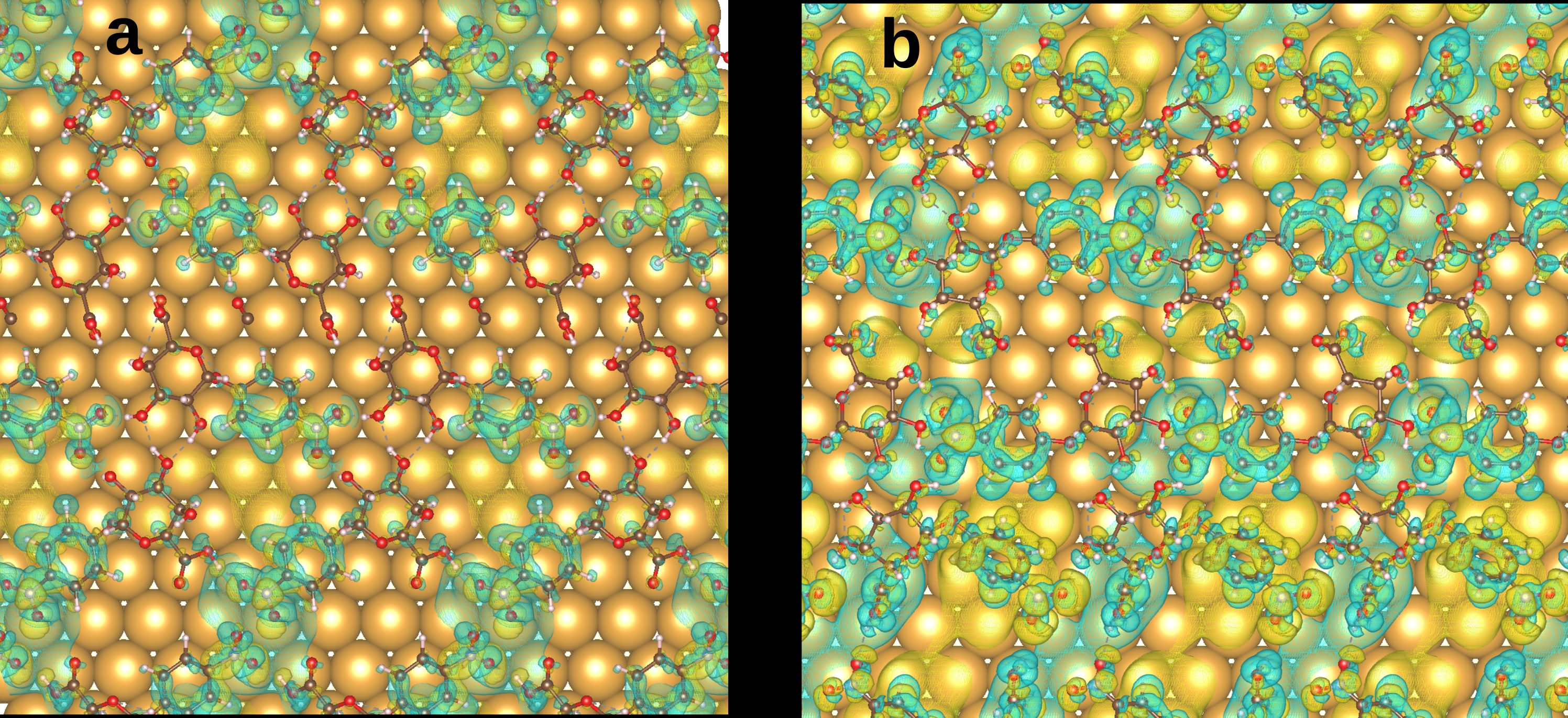}
    \caption{Charge density differences between the full system and the isolated substrate and molecules. for the $\alpha$ (a) and $\beta$ (b) self-assemblies computed with PBE+vdW$^{surf}$. Yellow isosurfaces correspond to positive electron density change (increased electron density), while blue correspond to negative changes (decreased electron density), isovalue = 0.0013.}
    \label{fig:charge_densities}
\end{figure}

According to the stereoelectronic effects, the more imbalanced charge distribution in the central conformations of both \textbf{NBDG} and \textbf{NADG} basic units suggests that they serve as stronger hydrogen bond donors and acceptors, forming stronger intermolecular bonds than the corresponding edge conformations. This rationalizes the presence of multiple conformations within the basic unit, as it enhances electron density delocalization, stabilizing the hydrogen-bonded structure. Furthermore, the enhanced electron density transfer from the \textit{n}${(O_1)}$ orbital into the aromatic nitrophenyl group in \textbf{NBDG} suggests increased rigidity of the molecule as seen through a more constant $\psi$ dihedral angle and a weaker interaction with the Au(111) surface compared to \textbf{NADG}. This is supported by a longer average distance between the nitrophenyl groups and the surface for the former at 3.84 Å, compared to 3.44 Å for the latter. The molecules are physisorbed onto the surface, where the additional negative charge donated to the nitrophenyl group reduces the van der Waals interactions mainly responsible for the adsorption\cite{zhang_insight_2017,sweet_insights_2024}. As shown in the side views (Figure~\ref{beta_AFM_STM}j and Figure~\ref{alpha_AFM_STM}j), the nitrophenyl ring in the $\alpha$-center lies mostly parallel to the surface, while in the $\beta$-edge, it is more tilted, indicating a stronger surface interaction in the $\alpha$-center, consistent with the electron density transfer trend. This is corroborrated by the computed charge density differences between the adsorbed system and the isolated monolayer and substrate, where the nitrophenyl group has clearly received more charge density in \textbf{NBDG} than in \textbf{NADG} (Figure \ref{fig:charge_densities}). Besides, the nitrophenyl ring in both central and edge conformations in \textbf{NADG} preferentially adsorb atop Au atoms, whereas \textbf{NBDG} shows no specific adsorption site preference.

\section{Conclusions}
Through our study, we have achieved the real-space observation of carbohydrate stereochemistry in supramolecular assemblies with atomic resolution and constructed consistent 3D models of these through a modeling protocol merging machine learning with well-established first-principles methods. Owing to these detailed structural models, we are able to link the structural and stereochemical properties of the individual carbohydrates to the on-surface hierarchical chiralities in their self-assembles. NBO analysis and charge density computations reveal that the stereochemistry of the glycosidic bond directly influences charge transfer to the aromatic nitrophenyl groups. This charge transfer modulates interactions between the nitrophenyl groups and the surface, competing with those between the former and the monosaccharide backbone, ultimately affecting other intermolecular interactions that are part of controlling the self-assembly process, in particular the hydrogen bonding. These competitive interactions control the adsorption selectivity of the individual carbohydrate molecules studied herein and contribute to chiral selection in the self-assembly. This finding provides new insights into chiral conferral in self-assembly and on-surface stereoselective catalysis of materials and compounds based on carbohydrates. Finally, we have elucidated the role of glycosidic bond stereochemistry in influencing the charge distribution of the groups attached to the non-anomeric carbon, an aspect less frequently studied compared to the well-understood anomeric effects.

\section{Methods}
\subsection{Sample preparation and ESD}

The Au(111) single crystal, obtained from MaTeck, was cleaned through repeated Ne$^+$ sputtering with a beam energy of 1000 eV and an ion current of 30 $\mu$A for 10 minutes, followed by annealing at approximately 450$^{\circ}$C for 5 minutes. Solutions of \textbf{NADG} and \textbf{NBDG} for ESD were prepared by dissolving the powders in a 1:1 (v/v) mixture of methanol and acetonitrile, resulting in a final concentration of $\sim$0.02 mmol/L. The flow rate during deposition was set to 955 $\mu$L/h. Positive ion mode was employed, and the voltage applied to the emitter ranged from $\sim$3000 to 3500 V over a duration of approximately 30 minutes. Throughout the deposition process, the sample was maintained at room temperature in an UHV chamber with a base pressure of $1 \times 10^{-9}$ mbar.

\subsection{STM/AFM measurements}
All experiments were conducted on a combined non-contact AFM/STM system (Createc) equipped with a commercial qPlus sensor with a Pt/Ir tip (resonance frequency $f_0 \approx 301470$ Hz and quality factor $Q \approx 77648$). The system operated with an oscillation amplitude of $A = 50$ pm at $T \approx 5$ K in UHV conditions, with a base pressure of approximately $1 \times 10^{-10}$ mbar.

\subsection{\textbf{DFT}}

DFT computations were done in FHI-AIMS \cite{Blum2009} and Gaussian \cite{g16}. The functional chosen for most of the computations was PBE augmented with the Tkatchenko-Scheffler dispersion correction parametrized for surfaces, termed PBE+vdW$^{surf}$ \cite{Tkatchenko2009,PhysRevLett.108.146103} with light defaults and first tier basis functions. We used a 1$\times$1$\times$1 Monkhorst-Pack grid to sample the Brilloun Zone. No spin polarization was employed due to the closed-shell character of the carbohydrates. The surface slab was constructed using four layers of 12$\times$4 Au in the fcc111 structure, where the two lowest layers were kept fixed. The unit cell dimensions were determined from the experimental line profiles, which were subsequently relaxed with DFT, resulting in monoclinic unit cells with dimensions (\textit{a}, \textit{b}, $\gamma$) = (11.71 Å, 29.96 Å, 111.1$^{\circ}$) for the \textbf{NADG} monolayer, (\textit{a}, \textit{b}, $\gamma$) = (11.70, 30.01, 73.18$^{\circ}$) for \textbf{NBDG}. With these lattice parameters, the underlying surface deviates from that of pristine Au(111), which could be attributed to variations in Au--Au distances resulting from the herringbone reconstruction of the real surface. In particular, we note that the relaxed model surface shows varying Au--Au distances both in the \textit{a} (4.97 - 5.00 Å) and \textit{b} (2.92 - 2.94 Å) directions of the face-centered unit cell, which is consistent with the experimentally observed anisotropic shortening of Au--Au distances \cite{tanaka_surface_2010}. To investigate donor-acceptor interactions, we used the Natural Bond Orbital analysis (NBO Version 3.1)\cite{glendening_nbo_nodate} as implemented in Gaussian. Here, as this analysis was done on the monolayers without including the surface, we opted for the B3LYP functional 
 \cite{PhysRevA.38.3098,PhysRevB.37.785} combined with the correlation-consistent cc-pVDZ basis set for a more accurate description of the electronic structure of the molecules. 

\subsection{\textbf{BOSS}}

 A quasi-random Sobol sequence was used to initialize the data, and we made use of the GP-Lower Confidence Bound acquisition function with increasing exploration (elcb). Standard periodic kernels were used for rotation and $xy$-translation, while radial basis functions (rbf) were used for the $z$-coordinate. Acquired data points were multiplied by leveraging surface symmetry, applying symmetry operations to the adsorbate at high-symmetry sites of the fcc111 surface. Convergence of the model was determined by checking if the predicted global minimum stopped changing between iterations. Initially, we conducted a search for the global minimum conformers of the two \textbf{NADG} and \textbf{NBDG} anomers, the degrees of freedom including the full rotation of the hydroxyl groups, carboxylate group, the two bonds involved in the glycosidic linkage as well as the $^1C_4$ to $^4C_1$ ring inversion (9D search), basing the surrogate model on 599 DFT data entries (DFT single point energies) for NADG, 405 for \textbf{NBDG}. We relaxed all of the predicted BOSS conformers, resulting in 83 and 107 unique conformers for \textbf{NADG} and \textbf{NBDG}, respectively. Following this, we used the global minimum conformer to search for the global adsorption minima. The surrogate model for the adsorption of single \textbf{NADG} molecules on the surface (6D search) was constructed out of 654 DFT data entries, while the \textbf{NBDG} counterpart was constructed out of 628 DFT data entries. Out of 244 unique \textbf{NADG} adsorption structures predicted by BOSS, we relaxed the lowest 10 using DFT. Similarly for the \textbf{NBDG} structures, the number of unique structures predicted were 185, and of these we relaxed 17 with DFT. We limited ourselves to a subset of the adsorption structures due to the high computational cost and time inherent to DFT, working under the assumption that the lowest energy predictions from BOSS would represent the lowest energy DFT-relaxed structures adequately. 

\subsection{\textbf{CREST}}
The Conformer-Rotamer Ensemble Sampling Tool (CREST, Version 2.12) \cite{pracht_automated_2020,pracht_crestprogram_2024} was employed as an auxiliary method to probe the accuracy of BOSS for the \textbf{NADG} conformer analysis. We were unable to use this to validate the adsorption structure search as it is currently unavailable for periodic systems. The length of the metadynamics run was determined to be 98 ps to fully sample the conformational phase space, based on a flexibility measure of 0.288. This resulted in 119 unique conformers, somewhat more than what was provided by BOSS, which is expected since CREST employs collective variables that allows for more flexibility in the ring configurations. Nonetheless, the two methods identify the same lowest energy conformer, providing support for the BOSS-based analysis.  

\subsection{\textbf{NequIP}}
We trained the MLIP Neural Equivariant Interatomic Potential (NequIP) \cite{batzner_e3-equivariant_2022} on DFT data to accelerate the screening of candidate monolayer structures. This represents the state-of-the-art within MLIPs, in particular with respect to its high data efficiency resulting from its E(3)-equivariant design. Nonetheless, it should be mentioned that an improved version --BOTNet-- has recently been described \cite{batatia2022designspacee3equivariantatomcentered}. The training data contained 2532 entries including the atomic coordinates, along with their computed DFT energies and forces, randomly split into 2232 for training and 300 for validation. All energies were modified by subtracting the atomization energies from the total system energies. The data included a mixture of the conformers of the isolated molecules, single molecule adsorbates, a selection of monolayer structures for both \textbf{NADG} and \textbf{NBDG}, and finally structures sampled from high-temperature NVE molecular dynamics (T = 4000 K) using the semi-empirical tight-binding DFT method GFN-xTB \cite{grimme_general_2017}. Both the trained NequIP model and GFN-xTB were used as calculators in the Atomic Simulation Environment (ASE) \cite{larsen_atomic_2017}. We used NequIP version 0.5.6 along with e3nn \cite{mario_geiger_2021_4735637} version 0.4.4. The cutoff radius was 4.5 Å, training batch size 5, validation batch size 10, while learning rate was set to 0.0075. The configuration file (\textit{config.yaml}) containing all training parameters, the trained model itself, as well as the training dataset can be found in the Zenodo repository (DOI:10.5281/zenodo.13990712).  

\subsection{\textbf{Minima hopping}}
The minima hopping algorithm was used with the trained NequIP model, employing Hookean bond constraints to preserve intact molecules during the high-temperature NVE MD propagation steps of the algorithm. The velocity verlet integrator was used to calculate the trajectories, the initial velocities were set by a Maxwell-Boltzmann distribution. The energy criteria (\textit{E$_{diff}$}) for accepting a new minimum was initially set to 2.5 eV. The initial temperature (\textit{T$_{0}$}) was set to 4000 K in an attempt to sample the configurational phase space efficiently. However, the molecular adsorbates are easily desorbed at such high temperatures, so we added a Hookean volatilization constraint, \textit{i.e.} a restorative force keeping the molecules from drifting further than 18 Å from the surface, since this height allows full rotation of the molecules on the surface. This procedure has been previously described in literature, and more details can be found in the original publication \cite{peterson_global_2014}. All minimum structures from the procedure can be found in the Zenodo repository (DOI:10.5281/zenodo.13990712). 

\subsection{\textbf{AFM simulations}}
The final AFM-images were simulated with the GPU-version of the Probe-Particle Model \cite{hapala_mechanism_2014,krejci_principles_2017} using the Hartree-potential of the sample and a \textit{dz$^2$}-multipole on the probe particle, mimicking a CO molecule at the apex of an SPM tip. The charge of the probe particle was set to \textminus0.05, determined by best apparent match with experiment. Exploration and adjustment of structures were done by simulating AFM images using only the geometry as input, providing similar images containing the same main features as with the more accurate simulations employing the Hartree-potential.  

\subsection{\textbf{STM simulations}}
STM images were simulated with FHI-aims, employing the Tersoff-Hamann approximation \cite{PhysRevB.31.805} with bias 0.2 V. Visualization was done with Vesta and the WSxM software \cite{10.1063/1.2432410} with isovalues typically kept in the range of 10$^{-10}$ to 10$^{-12}$ based on match with experiment.

\section*{Acknowledgements}
The authors thank the Centre for Scientific Computing (CSC) for resources in projects ay6310 and 2008059, as well as the Aalto Science IT project
for computational resources. We also thank the Academy of Finland (AoF) for grant number 347319 funding the project “Microscopy and machine
learning in molecular characterization of lignocellulosic materials” (MIMIC). This work was undertaken as part of the FinnCERES competence centre.

\section*{Author contributions}
S.C. conducted the SPM experiments. J.S.J. assisted in the experiments, trained the MLIP, and implemented the minima hopping and DFT calculations. S.C. and J.S.J. prepared the manuscript with input from all authors.

\section*{Supporting information}
The supporting information contains supplementary experimental data in the form of large-scale STM images and topography line profiles. Moreover, it contains supplementary computational data from Bayesian Optimization Structure Search (BOSS), details from the minima hopping procedure, simulated large-scale AFM images and their overlap with experiment, comparison of experimental and simulated STM and AFM images for a subset of candidate structures, analysis of the rotational barriers for hydroxyl and carboxylate groups in the NADG assembly, as well as a summary of the orbital interactions relevant to the donor-acceptor orbital interactions in the asssemblies. 

Additional computational data including conformers, single adsorbate, and monolayer structures, along with trained NequIP potentials and data can be found in Zenodo repository (\url{DOI:10.5281/zenodo.13990712}).

\newpage

\tableofcontents


\clearpage
\addcontentsline{toc}{section}{Experimental}
\section*{Experimental}
\addcontentsline{toc}{subsection}{Supplementary experimental data: large-scale STM images}
\subsection*{Supplementary experimental data: large-scale STM images}\label{sec:exp}

\clearpage
\addcontentsline{toc}{subsection}{Supplementary experimental data: STM topography line profiles}
\subsection*{Supplementary experimental data: STM topography line profiles}\label{sec:line_prof}

\begin{figure}[htbp]%
    \centering
    \includegraphics[width=\textwidth]{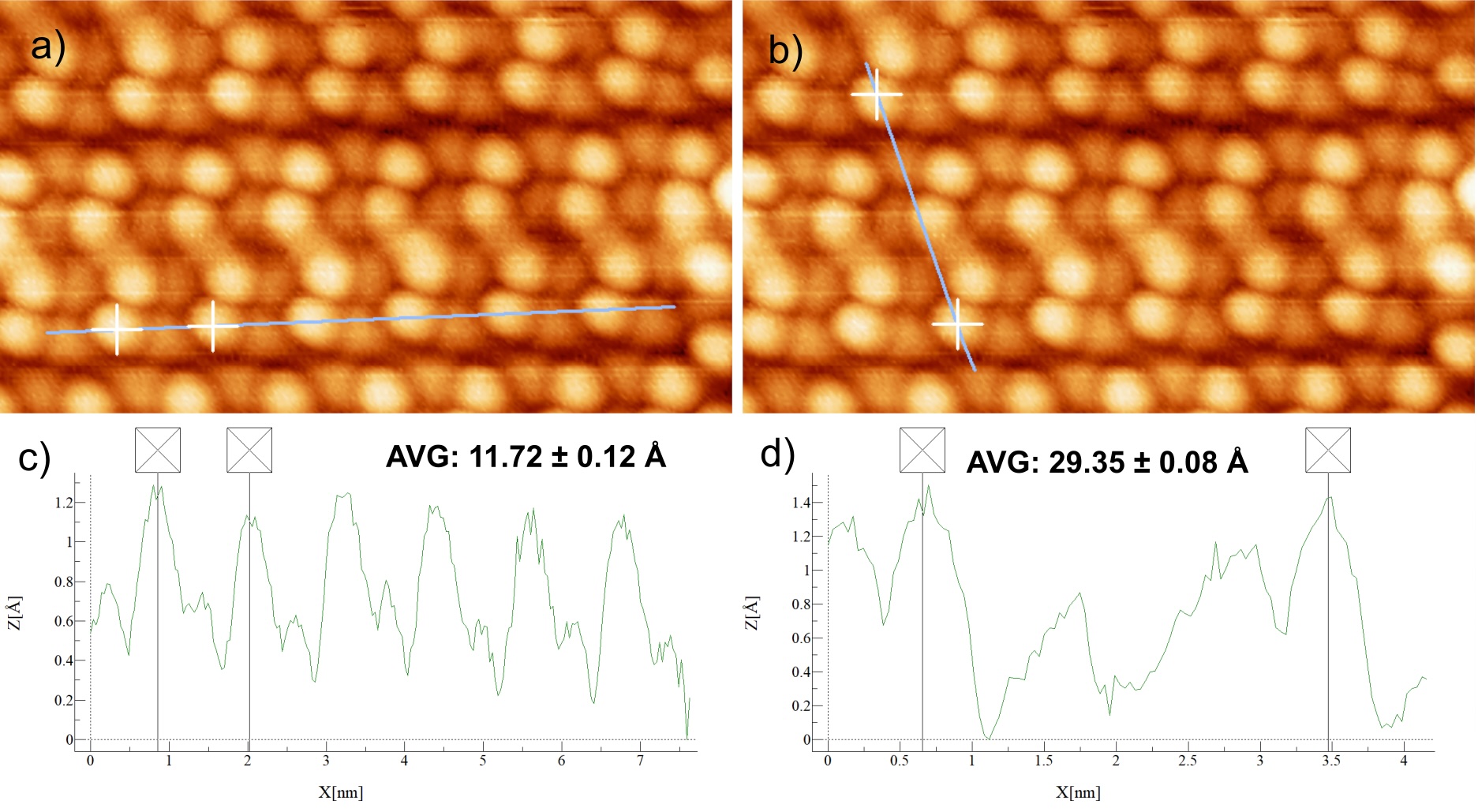}
    \caption{Determination of lattice parameters for the $\alpha$-monolayer from STM line profiles. a) and c) shows the line profile for the \textit{a}-vector, b) and d) shows the line profile for the \textit{b}-vector. The angle between the two ($\gamma$) was found to be 107$^{\circ}$.}
    \label{fig:alpha_lines}
\end{figure}

\clearpage

\begin{figure}[htbp]%
    \centering
    \includegraphics[width=\textwidth]{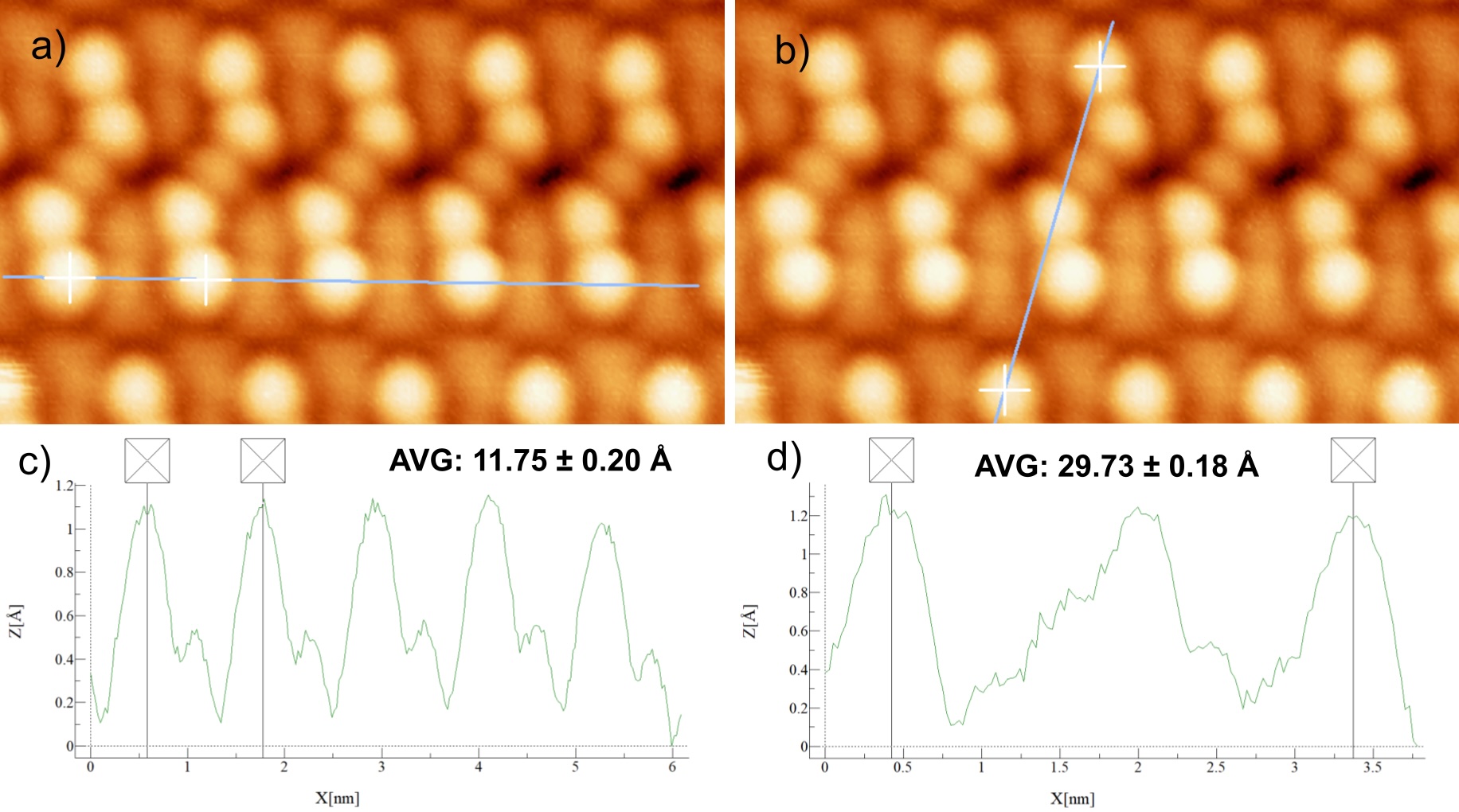}
    \caption{Determination of lattice parameters for the $\beta$-monolayer from STM line profiles. a) and c) shows the line profile for the \textit{a}-vector, b) and d) shows the line profile for the \textit{b}-vector. The angle between the two ($\gamma$) was found to be 75$^{\circ}$.}
    \label{fig:beta_lines}
\end{figure}

The dimensions of the repeating units for NADG and NBDG self-assemblies were determined by averaging five repeating units in both the a- and b- directions as illustrated in Figures \ref{fig:alpha_lines} and \ref{fig:beta_lines}, the average dimensions for NADG being (a, b, $\gamma$) = (11.72 Å, 29.35 Å, 107$^{\circ}$), the NBDG counterparts (\textit{a}, \textit{b}) = (11.75 Å, 29.73 Å, 75$^{\circ}$), in fair agreement with the DFT-relaxed unit cell dimensions. However, the experimental \textit{b}-vector is consistently smaller than the theoretical counterpart by 0.61 and 0.28 Å for NADG and NBDG.

\clearpage

\begin{figure}[htbp]%
    \centering
    \includegraphics[width=\textwidth]{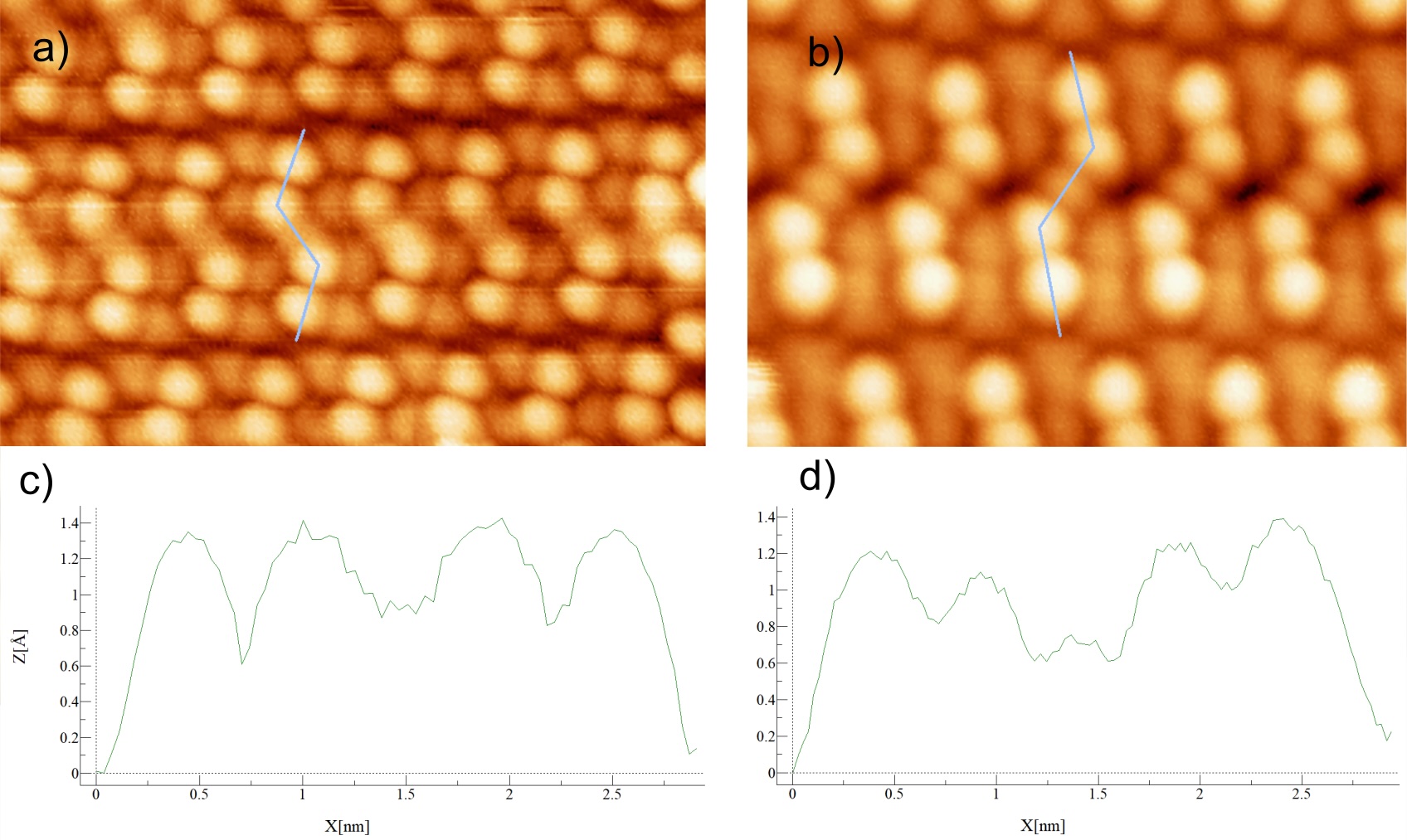}
    \caption{Line profiles along the highest parts of the repeating units for $\alpha$- (a, c) and $\beta$-monolayers (b, d). The profiles are drawn from top to bottom based on the a) and b) images.}
    \label{fig:alpha_beta_lines}
\end{figure}

To gain further insight into the detailed structures of the monolayers, we looked at the line profiles across the parts that were protruding the furthest from the surface as illustrated in Figure \ref{fig:alpha_beta_lines}. Both NADG and NBDG display similar profiles, with four higher protrusions in addition to a sligthly lower one corresponding to the area in between the two central units. However, we note one minor difference between the two assemblies, where the central units in NADG are slightly higher, or at similar heights than the edges, while the opposite is true for NBDG. These observations are consistent with the AFM images, with the center of NADG showing up as brighter features earlier in the AFM image height stack, while the NBDG structure display more sharp features at the edges than in the center.  

\clearpage
\addcontentsline{toc}{section}{Computational}
\section*{Computational}
\addcontentsline{toc}{subsection}{Bayesian Optimization Structure Search (BOSS)}
\subsection{Bayesian Optimization Structure Search (BOSS)} 

\begin{figure}[htbp]%
    \centering
    \includegraphics[width=\textwidth]{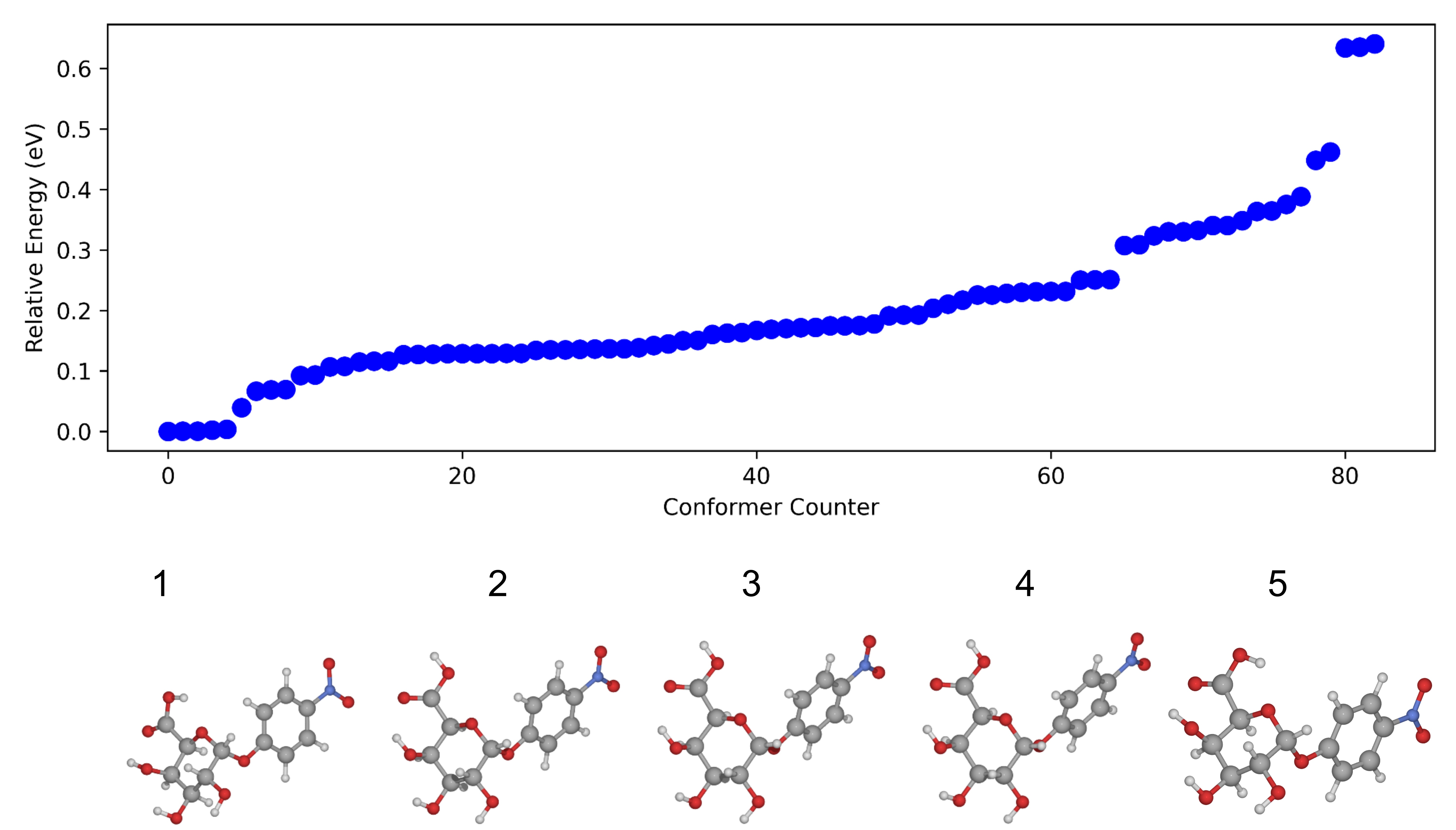}
    \caption{PBE+vdW$^{surf}$ relaxed conformer energies and the five lowest energy $\alpha$-anomer structures predicted by BOSS.}
    \label{fig:BOSS_isolated_molecules_alpha}
\end{figure}
Figure \ref{fig:BOSS_isolated_molecules_alpha} displays the results of the conformational analysis for the $\alpha$-anomer of 4-nitrophenyl-D-galacturonide. The lowest energy structure has a clockwise H-bond network involving the -OH and -COOH groups the of the galacturonide moiety. While the hydrogen in the -COOH group is here in a \textit{trans}-orientation towards the O$_5$-atom, an isoenergetic conformer has it in the \textit{cis}-orientation, i.e. towards the carbonyl oxygen. A final intramolecular H-bond is formed between the O$_5$-atom and a ortho-hydrogen in the nitrophenyl group, located above the central galacturonide plane. The rest of the structures differ in the rotation of the nitrophenyl ring with respect to the galacturonide moiety.    

\clearpage

\begin{figure}[htbp]%
    \centering
    \includegraphics[width=\textwidth]{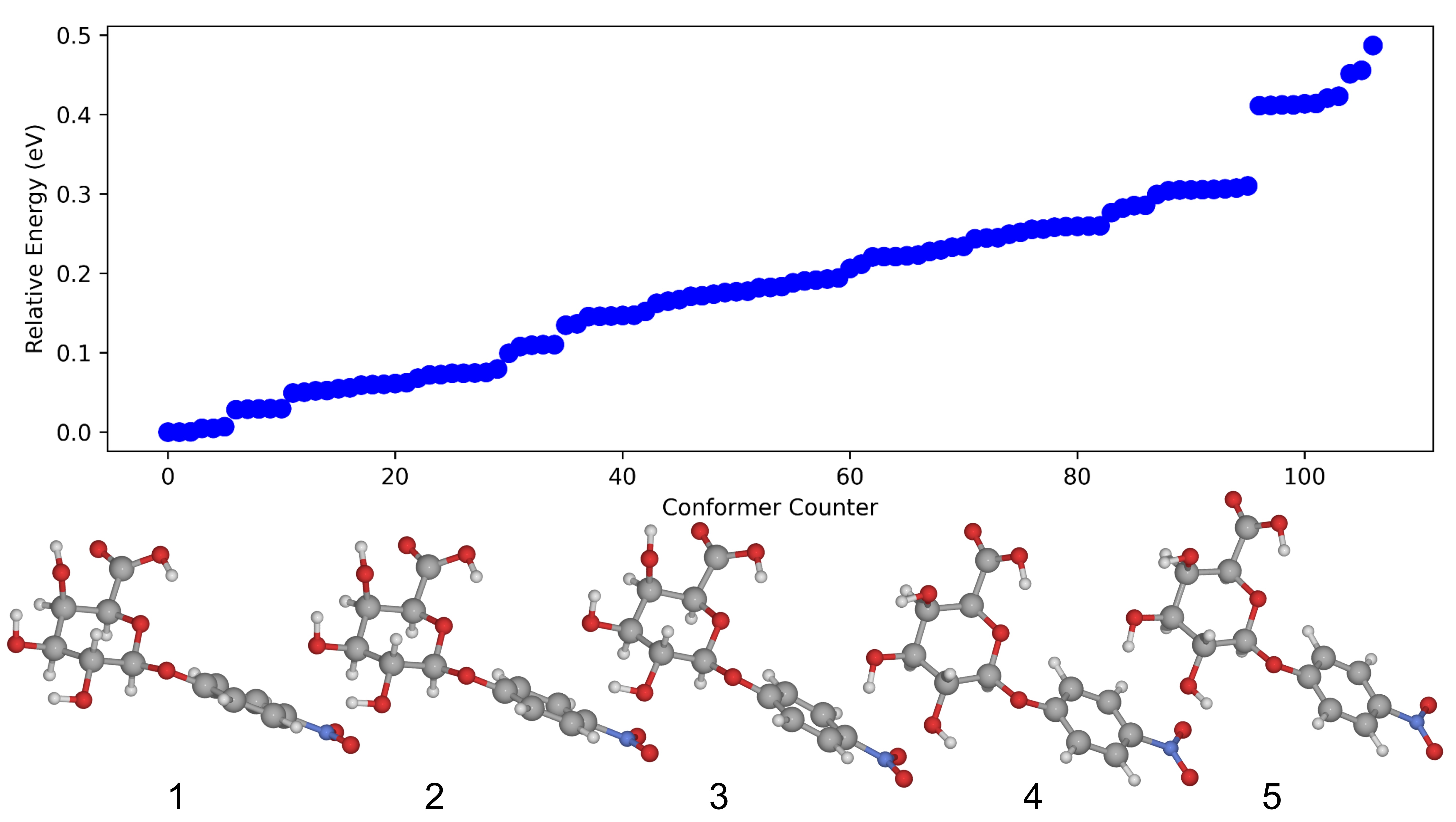}
    \caption{PBE+vdW$^{surf}$ relaxed conformer energies and the five lowest energy $\beta$-anomer structures predicted by BOSS.}
    \label{fig:BOSS_isolated_molecules_beta}
\end{figure}

Figure \ref{fig:BOSS_isolated_molecules_beta} displays the results of the conformational analysis for the $\beta$-anomer of 4-nitrophenyl-D-galacturonide. The structures determined are similar to those for the corresponding $\alpha$-anomer, with a similar clockwise H-bond network. However, due to the equatorial placement of the nitrophenyl group, the orientation of the latter makes the H-bond to the O$_5$-atom now appear below the central galacturonide plane. This is likely due to the above-plane alternative bringing the hydrogen on the nitrophenyl group too close to that of the COOH-group, with added repulsion compared to the below-plane bond. Furthermore, the fourth lowest $\beta$-anomer displays a counter-clockwise H-bond network 0.005 eV (0.5 kJ/mol) higher than the global minimum, while the $\alpha$-counterpart has this as the sixth in line, 0.04 eV (4 kJ/mol) higher than the global minimum. 

\clearpage

\begin{figure}[htbp]%
    \centering
    \includegraphics[width=0.8\textwidth]{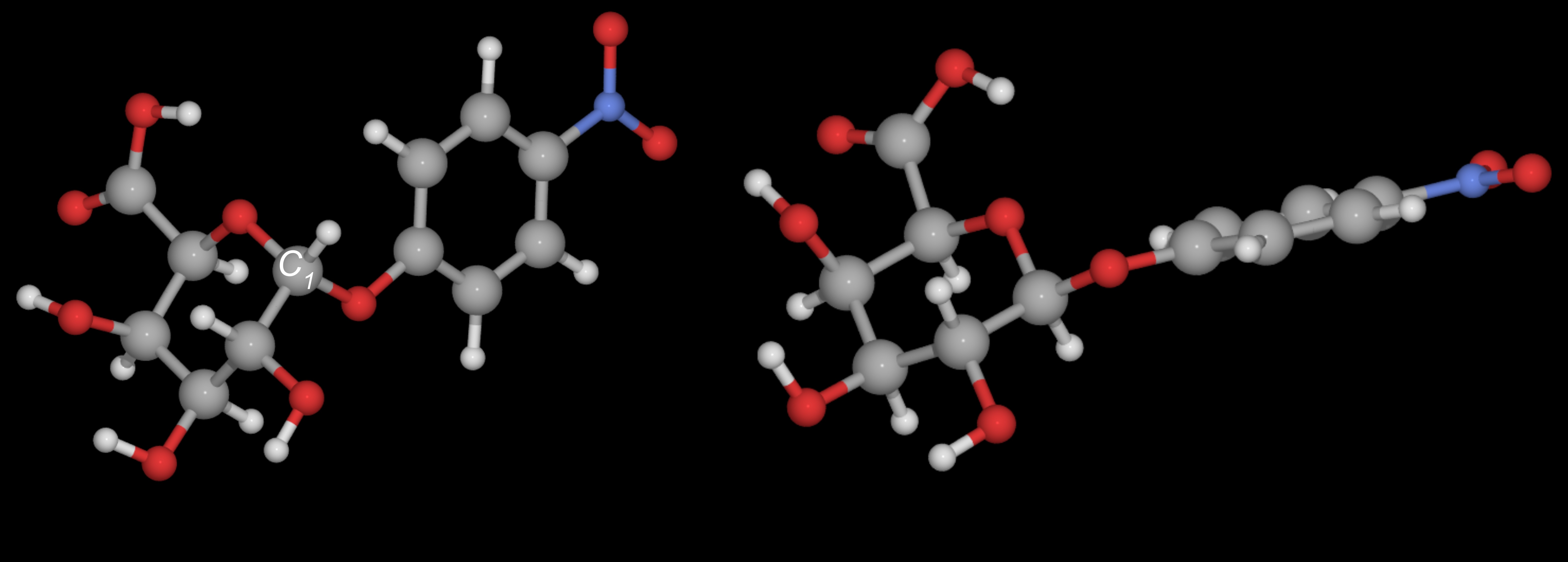}
    \caption{Global minimum conformers for isolated $\alpha$- and $\beta$-anomers identified using BOSS.  Relevant bond distances for the anomeric effect are shown below each molecule.}
    \label{fig:isolated_molecules}
\end{figure}

By analyzing the donor-acceptor interactions \textit{via} second order perturbation theory in the natural bond orbital (NBO) basis\cite{weinhold_what_2016} for the global minimum conformers, we note that the endo and exo anomeric effects are balanced in the $\alpha$-anomer, both providing around 50 kJ/mol stabilization, leading to nearly equal C$_{1}$--O$_{5}$ and C$_{1}$--O$_{1}$ bonds at 1.43 and 1.42 Å. Meanwhile, for the $\beta$-anomer, the exo anomeric effect is stronger than the endo anomeric effect, providing 58 and 16 kJ/mol of stabilization, respectively. This implies that for the latter, donation of the exocyclic oxygen lone pair (\textit{n}$_{(O1)}$) to the $\sigma$* orbital of the C$_{1}$--O$_{5}$ bond is more efficient than donation of the endocyclic oxygen lone pair (\textit{n}$_{(O5)}$) to the C$_{1}$--O$_{1}$ $\sigma$* orbital, which in turn is reflected in the corresponding bond distances at 1.44 and 1.40 Å, respectively. In addition, the difference in energy for the two anomers is computed to be 7 kJ/mol (PBE+vdW$^{surf}$), with the $\alpha$-anomer being lowest, which is consistent with the anomeric effect stabilizing the $\alpha$-anomer. 

\clearpage

\begin{figure}[htbp]%
    \centering
    \includegraphics[width=\textwidth]{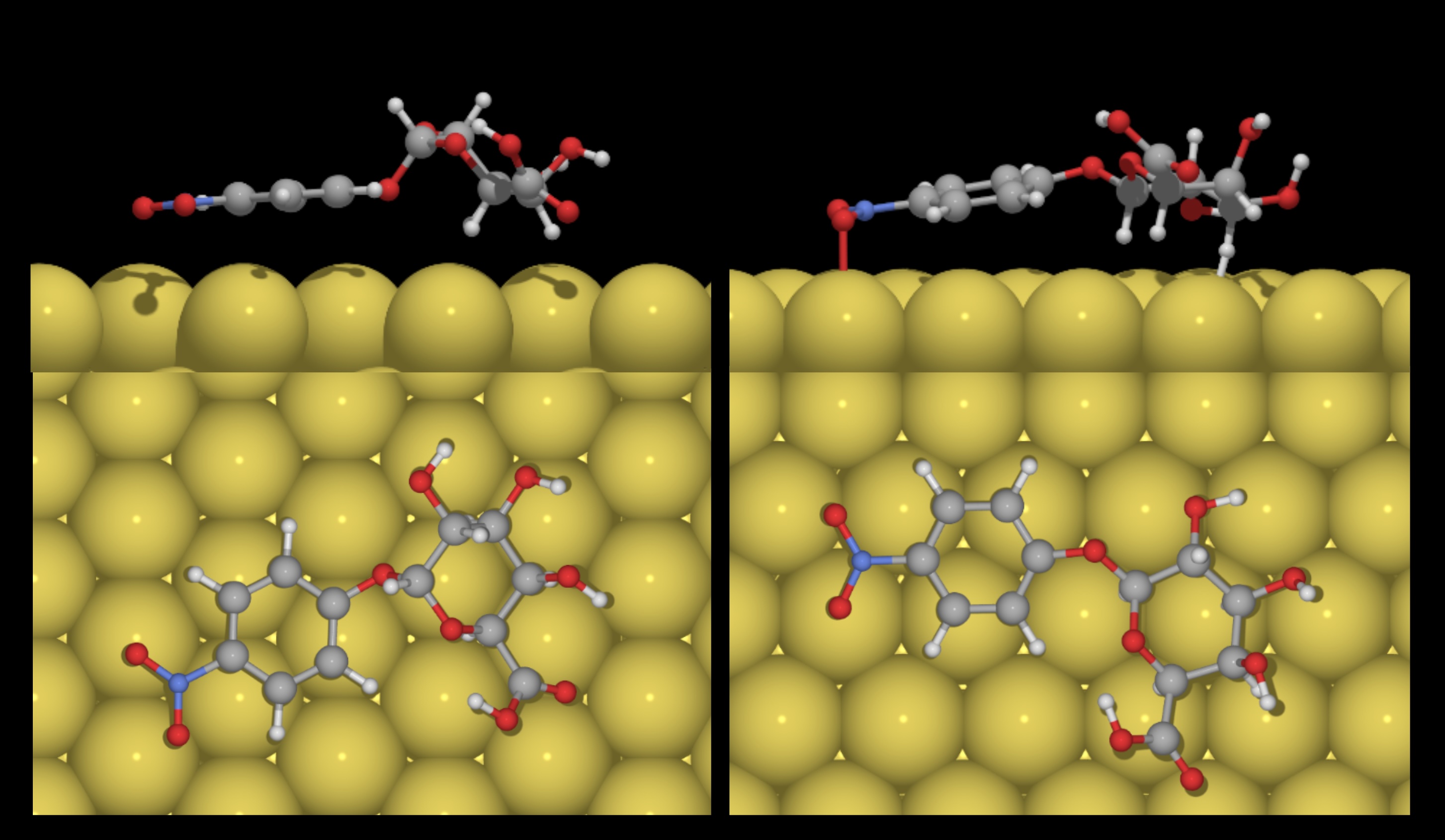}
    \caption{Global minimum PBE+vdW$^{surf}$-relaxed adsorption structures for single $\alpha$- and $\beta$-anomer molecules as identified by BOSS.}
    \label{fig:isolated_molecules_on_surface}
\end{figure}

Following the conformer analysis, the same approach was used to find the most stable adsorption configuration of single NADG and NBDG adsorbates, this time employing the global minimum isolated molecules as building blocks in the adsorption structure search, while the surface slab was acting as the second building block for BOSS. The surface slab for the single molecule adsorbates was constructed from four layers of 7$\times$8 Au atoms in the fcc111 crystal structure. The most stable adsorption configurations of the single molecules determined by BOSS differ in that the glycosidic bond is pointing towards the surface in NADG, while it is pointed away from the surface in NBDG (Figure \ref{fig:isolated_molecules_on_surface}). The adsorption energies of the two are similar at \textminus1.60 eV (\textminus154.2 kJ/mol) for NADG, \textminus1.65 eV (\textminus159.5 kJ/mol) for NBDG, the difference being close to the magnitude of the anomeric effect stabilizing the NADG molecule. Meanwhile, flipping the NBDG molecule to orient its glycosidic bond towards the surface leads to an adsorption configuration only 0.10 eV (9.2 kJ/mol) higher in energy than the global minimum. Similarly, flipping over NADG costs 0.11 eV (11.5 kJ/mol). Considering the multiple possible hydrogen bond donors and acceptors on each molecule, and the typical H--O$\cdots$H--O bond energy being around 0.22 eV (21 kJ/mol), the global minimum configurations of the single adsorbates are not necessary equivalent to those attained by the molecules in monolayer as a more favorable hydrogen bonding pattern might counterbalance the lower energy of a given rotational state of the molecule on the surface. 

\clearpage
\addcontentsline{toc}{subsection}{Minima hopping for monolayer structures}
\subsection{Minima hopping for monolayer structures}

\begin{figure}[htbp]
    \centering
    \includegraphics[width=\textwidth]{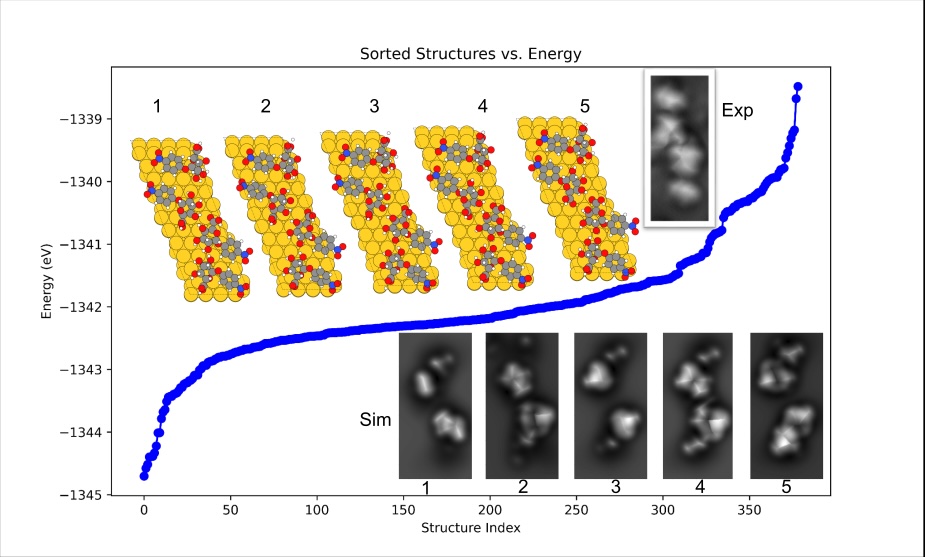}
    \caption{Global minimum monolayer structures for the $\alpha$-anomer from minima hopping using NequIP and their simulated AFM images. Numbering indicates the relative energy of the structure, with increasing number corresponding to higher energy. Experimental AFM image shown as inset.}
    \label{fig:minima_alpha}
\end{figure}

\clearpage

\begin{figure}[htbp]
    \centering
    \includegraphics[width=\textwidth]{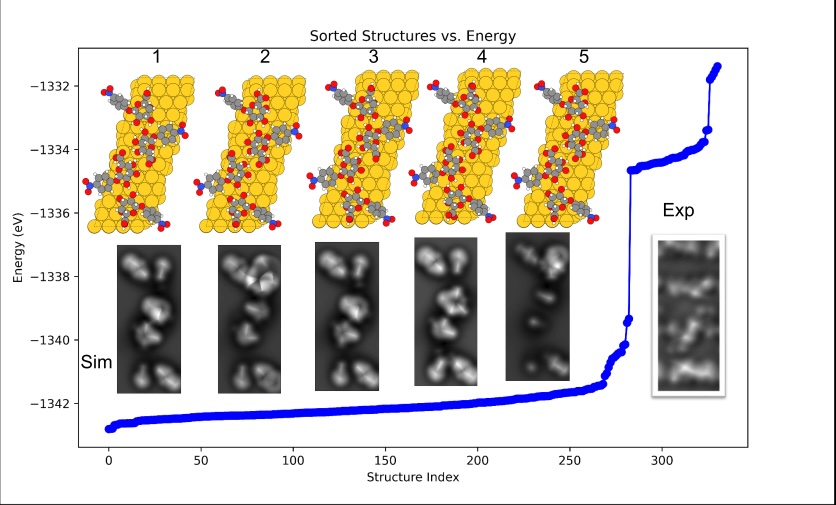}
    \caption{Global minimum monolayer structures for the $\beta$-anomer from minima hopping using NequIP and their simulated AFM images. Numbering indicates the relative energy of the structure, with increasing number corresponding to higher energy. Experimental AFM image shown as inset.}
    \label{fig:minima_beta}
\end{figure}

Initial monolayer structures were subsequently constructed by hand using the most stable single molecule adsorption configurations as building blocks, closely guided by the experimental constant current STM images. Subsequently, we simulated the AFM images of the initial 50 structures with the Probe-Particle Model,\cite{hapala_mechanism_2014,krejci_principles_2017}. Here it became apparent that the structures were merely close enough to the experiment to recognize single features in the images. Furthermore, manual construction of the monolayer structures and the following DFT relaxation was a highly time-consuming and computationally expensive endeavor. To accelerate the screening of the required configurations, we trained an MLIP, NequIP,\cite{batzner_e3-equivariant_2022} on the already acquired data, supplemented by high-temperature molecular dynamics snapshots from the semiempirical tight-binding DFT method GFN-xTB\cite{grimme_general_2017}, subsequently recomputed using the same DFT functional (PBE+vdW$^{surf}$) as the rest of the data entries. The purpose of the latter was to improve the reliability of the MLIP by including a more complete configurational phase space in the training data. With the trained MLIP, we used the minima hopping algorithm as outlined by Goedecker\cite{goedecker_minima_2004}, placing Hookean constraints the adsorbates to preserve molecular identity and to restrict them from sampling irrelevant parts of the phase space, such as the vacuum region\cite{peterson_global_2014}. Minima hopping was chosen here as BOSS is restricted to less than 20 degrees of freedom, which is sufficient for the conformer analysis and the isolated adsorbates, yet insufficient to establish the structures of the full monolayers. This sampling provided 850 monolayer structures in total for the two anomers, and along with them rough versions of the experimentally observed monolayer structures. 

The global minimum structures provided by the minima hopping procedure are overall close to those imaged experimentally--in particular for NBDG--but with minor discrepancies in functional group orientations and in the heights of specific atoms, which leads to noticeable differences in the simulated images. The final structures leading to the closest apparent match with experimental images were determined by altering the positions of mismatching atoms by hand, while fixing the atoms responsible for contrast similar to experiment. These altered structures were pre-relaxed with the MLIP before final DFT relaxations to reduce the amount of time spent on the latter. It should be noted that the final structures were kept partially constrained during relaxation. Specifically, the atoms surrounding the carboxyl groups of the central monosaccharide units of both structures were fixed in a position higher than in the relaxed counterparts. This suggests that there could be some feature in the surface which we were unable to capture with our methods. This could for instance be adatoms, an assumption supported by the fact that the monolayers are formed close to step edges, potentially acting as a source of adatoms. It has also been shown that organic molecules deposited onto a surface with solvent, such as methanol, may expedite adatom migration relative to UHV deposition, consistent with the above hypothesis\cite{ghalgaoui_nanostructuring_2017}. 
A second possibility is that the electrospray deposition process itself introduces either solvent molecules or contaminants which could be incorporated into the monolayer structure, such as Na or Ca ions, known to leach from laboratory glassware\cite{ledieu_leaching_2004}. A final third option could be that the slightly elongated units cells of our model structures do not fully capture the packing effects due to neighbouring adsorbates, with the real systems being slightly more crowded than our models. 

\begin{figure}[htbp]
    \centering
    \includegraphics[width=\textwidth]{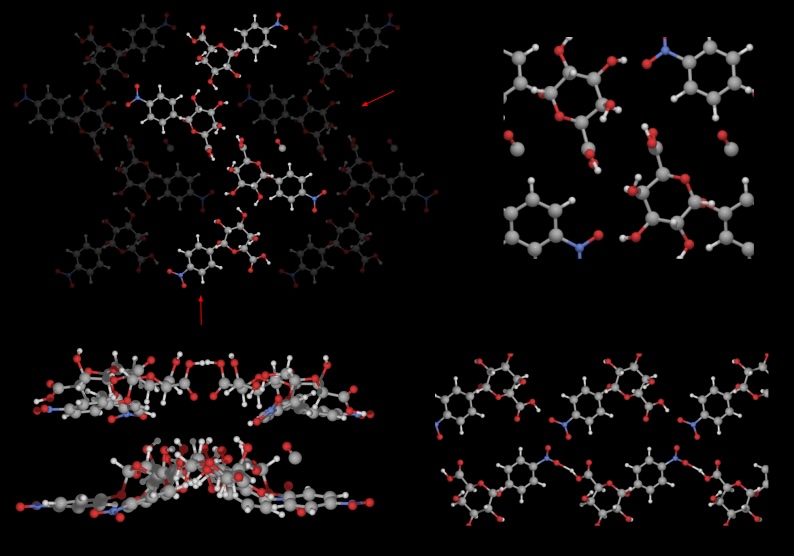}
    \caption{Final structural model of the NADG assembly.}
    \label{fig:final_alpha_assembly}
\end{figure}

\begin{figure}[htbp]
    \centering
    \includegraphics[width=0.8\textwidth]{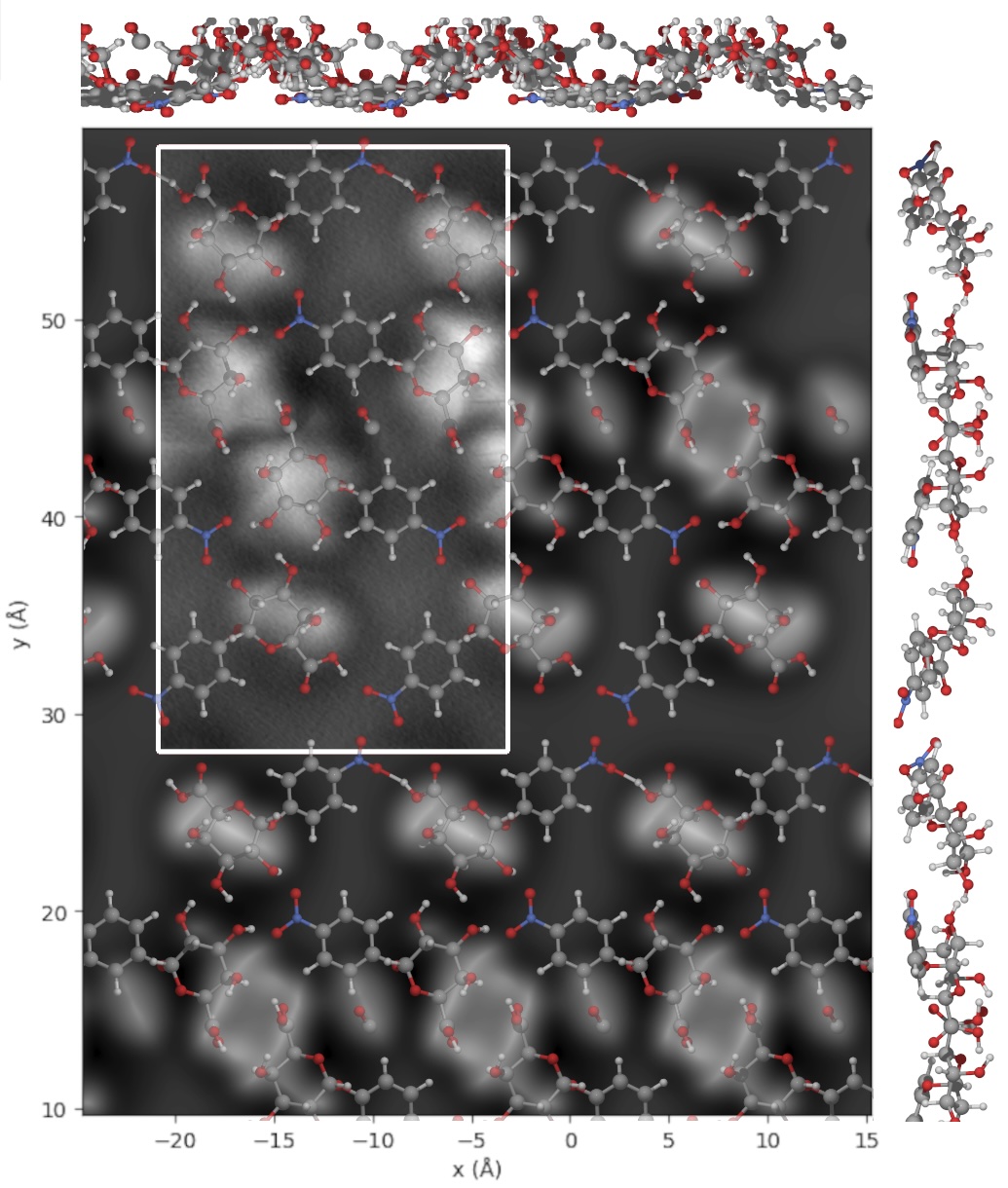}
    \caption{Comparison of simulated (larger image) and experimental (inset) AFM images with superimposed molecular structures for the $\alpha$-anomer.}
    \label{fig:AFM_large_scale_alpha}
\end{figure}

Although we initially assumed that the interactions between pairs of carboxyl groups would be a key driving force in the self assembly process, our structural models contradicts this. Specifically, a highly symmetric configuration with two O--H$\cdots$O=C contacts --akin to the most stable formic acid dimer-- is completely absent here, which we assume to be closely related to the most efficient packing of the adsorbates on the surface. However, we do observe a pair of closely related asymmetric structural motifs in the center of NADG, with one O--H$\cdots$O--H and one C=O$\cdots$H--C contact. These results can be rationalized using the formic acid dimer as a prototypical model of the interactions of the carboxyl groups in our systems, noting how the latter is analogous to the fourth most stable formic acid dimer structure\cite{farfan_dimers_2017, rodziewicz_formic_2009}. Based on accurate quantum chemical computations\cite{farfan_dimers_2017}, the interaction energy of the symmetric formic acid dimer is 53.7 kJ/mol, while that of the asymmetric motif is 23.3 kJ/mol. Hence, forming two of the latter does not quite balance out the stronger interaction of the symmetric structure in this model system. However, as shown in the structural model of the NADG monolayer in Figure \ref{fig:final_alpha_assembly}, this asymmetric bonding motif involves the hydroxymethyl -OH groups in a cyclic hydrogen bond network with the -OH groups of the $\alpha$-carboxyl, which is likely strengthened by co-operativity effects\cite{fonseca_guerra_telomere_2011,nochebuena_origin_2017,trevisan_quantitative_2022}. This cyclic network is directly observable in the AFM images, responsible for the contrast of the central features in the far and mid tip-distance images shown in Figure \ref{fig:AFM_large_scale_alpha}. In addition to these local effects, the asymmetric structure allows a more close packing of the central adsorbates, which extends to the full monolayer. 

In the final NADG structural model, a CO molecule was included between the 1D central units in the propagation direction to account for this seemingly isolated feature in the experimental AFM image. We note that this is not present between each repeating unit. 

\clearpage
\addcontentsline{toc}{subsection}{Large-scale AFM overlap}
\subsection{Large-scale AFM overlap}

\begin{figure}[htbp]
    \centering
    \includegraphics[width=0.8\textwidth]{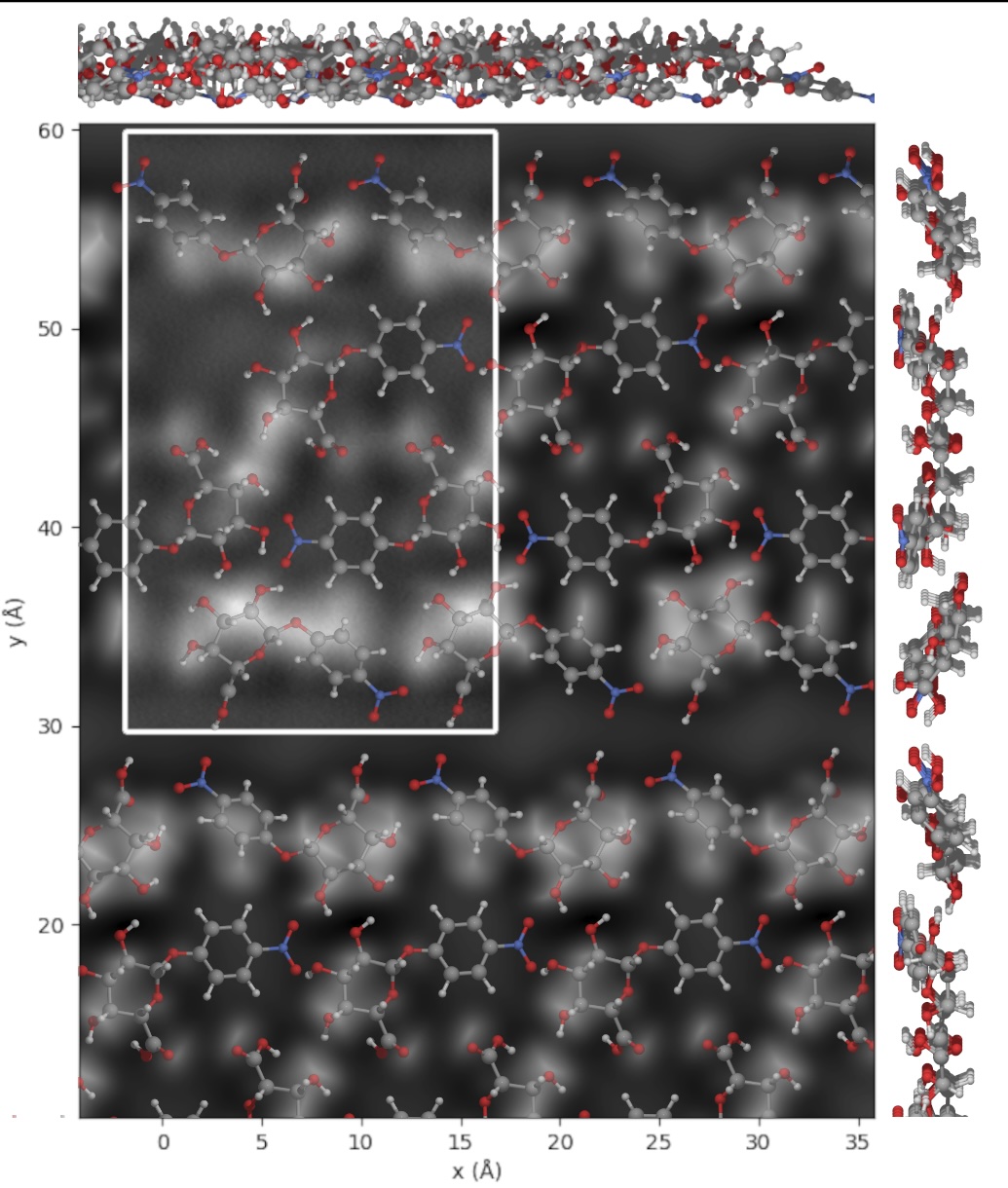}
    \caption{Comparison of simulated (larger image) and experimental (inset) AFM images with superimposed molecular structures for the $\beta$-anomer.}
    \label{fig:AFM_large_scale_beta}
\end{figure}

\clearpage
\addcontentsline{toc}{subsection}{Comparison of experimental and simulated STM and AFM images}
\subsection{Comparison of experimental and simulated STM and AFM images}

\begin{figure}[htbp]
    \centering
    \includegraphics[width=0.9\textwidth]{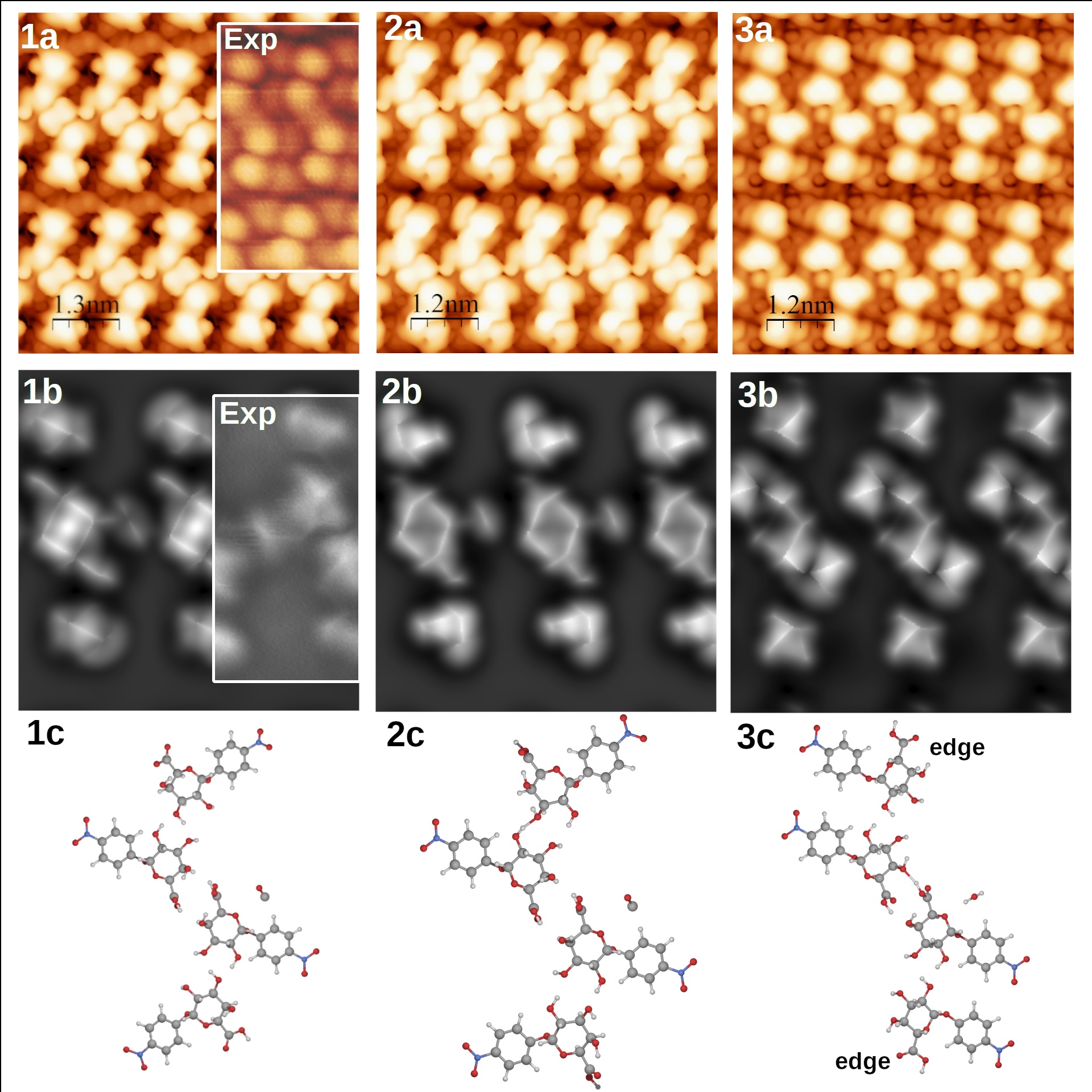}
    \caption{A selection of $\alpha$-monolayer structures (1c-3c) and their simulated STM (a) and AFM (b) images. Experimental STM and AFM images are shown in 1a and 1b as insets.}
    \label{fig:similarities_alpha}
\end{figure}

\begin{figure}[htbp]
    \centering
    \includegraphics[width=0.9\textwidth]{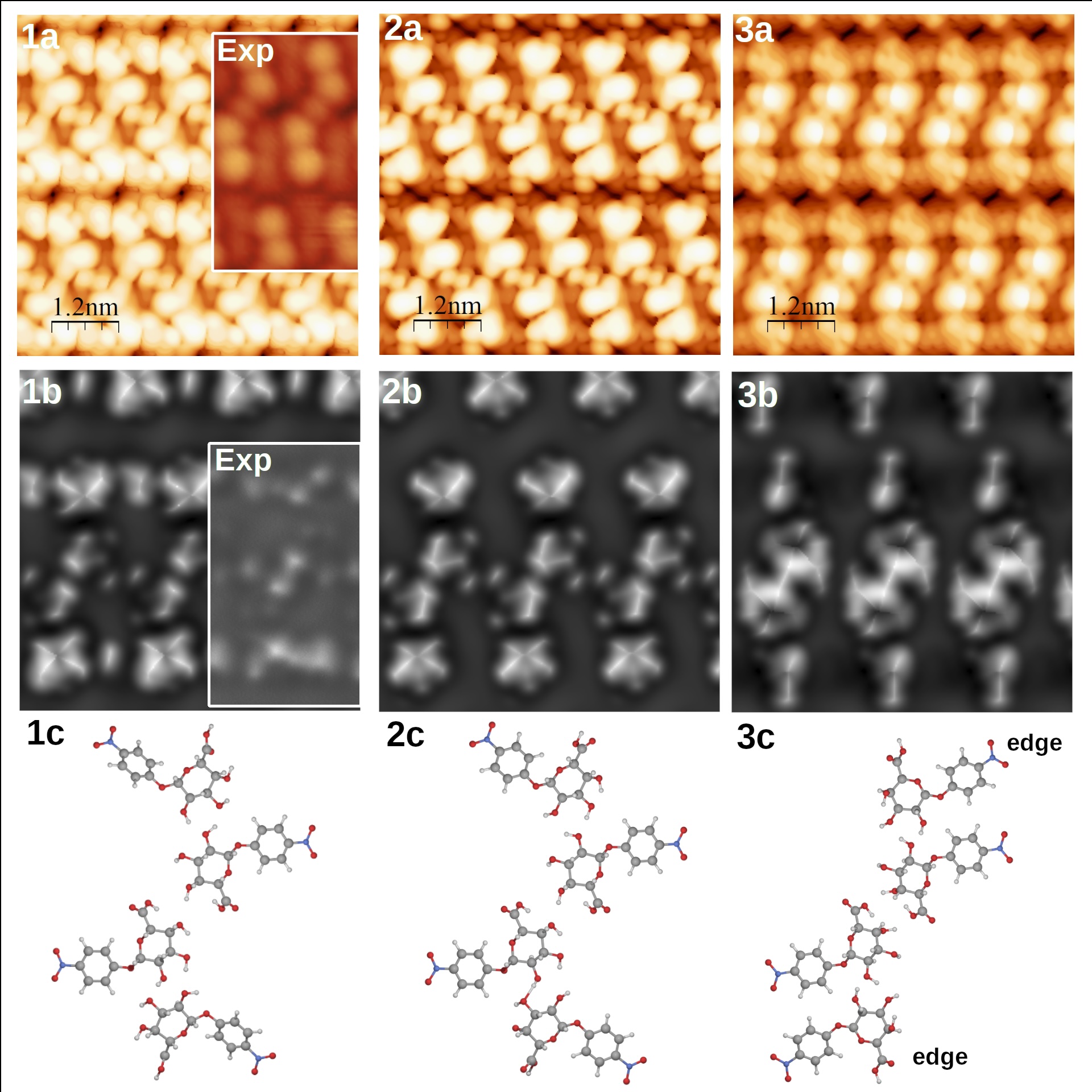}
    \caption{A selection of $\beta$-monolayer structures (1c-3c) and their simulated STM (a) and AFM (b) images. Experimental STM and AFM images are shown in 1a and 1b as insets.}
    \label{fig:similarities_beta}
\end{figure}

\clearpage

\addcontentsline{toc}{subsection}{Rotational barriers of hydroxyl and carboxylate groups in the NADG assembly}
\subsection{Rotational barriers of hydroxyl and carboxylate groups in the NADG assembly} 

\begin{figure}[htbp]%
    \centering
    \includegraphics[width=\textwidth]{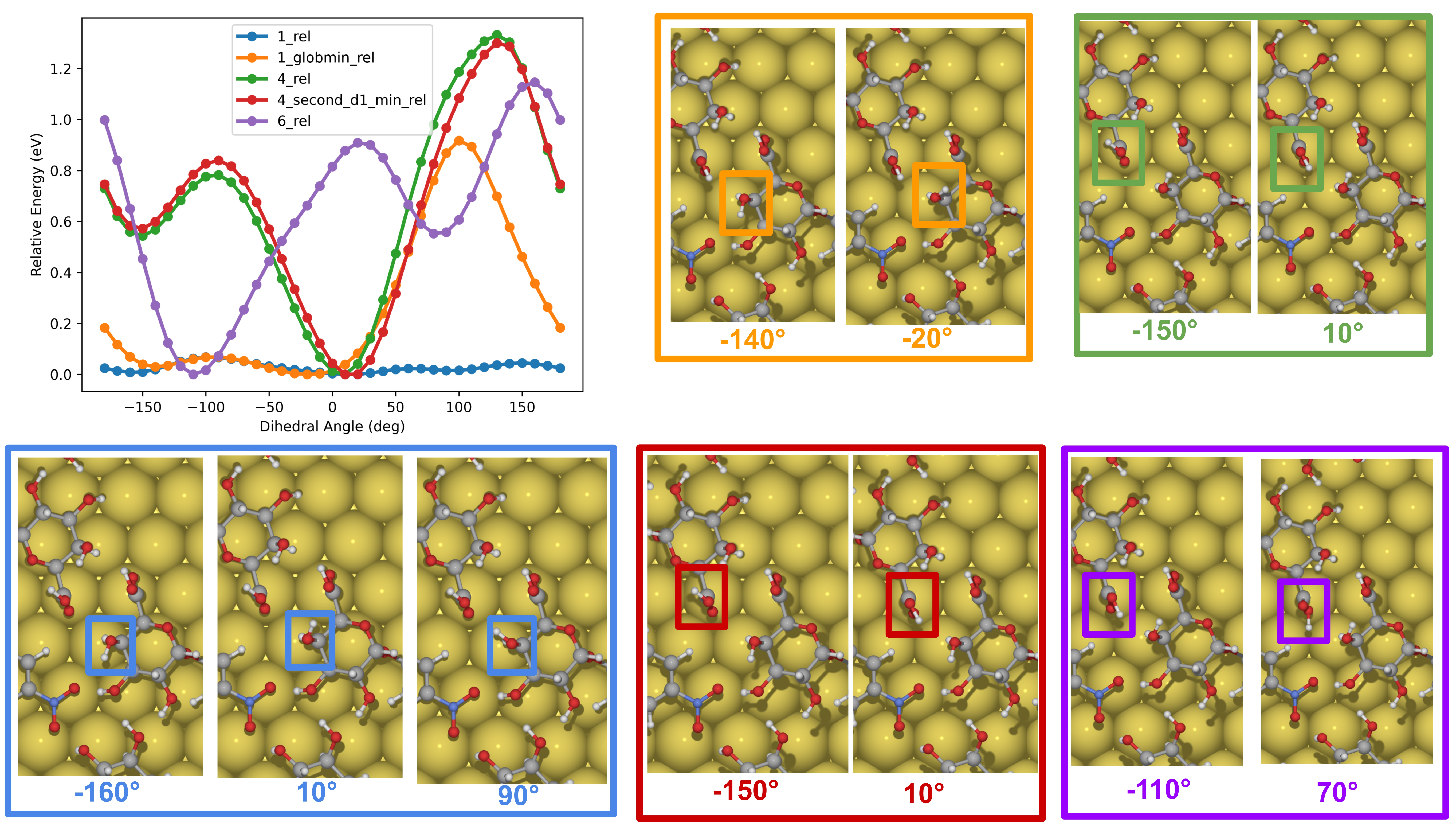}
    \caption{Energy curves for -OH rotation in the center of the NADG assembly. Orange marks the rotation of the bottom/top -OH groups (d1) with the dihedral angles of the right/left -OH groups kept identical to the experimentally determined model structure. Blue marks the rotation of bottom/top -OH groups (d4) with d1 = \textminus150$^{\circ}$. Green marks the rotation of the right/left -OH groups with the bottom/top groups kept identical to the experimentally determined model structure. Red marks the rotation of the right/left -OH groups with d1 = \textminus140$^{\circ}$. Purple marks the rotation of the right/left -COOH groups (d6) with all -OH groups kept identical to the experimentally determined model structure.}
    \label{fig:alpha_OH_rotational_energy_analysis}
\end{figure}

\begin{figure}[htbp]%
    \centering
    \includegraphics[width=\textwidth]{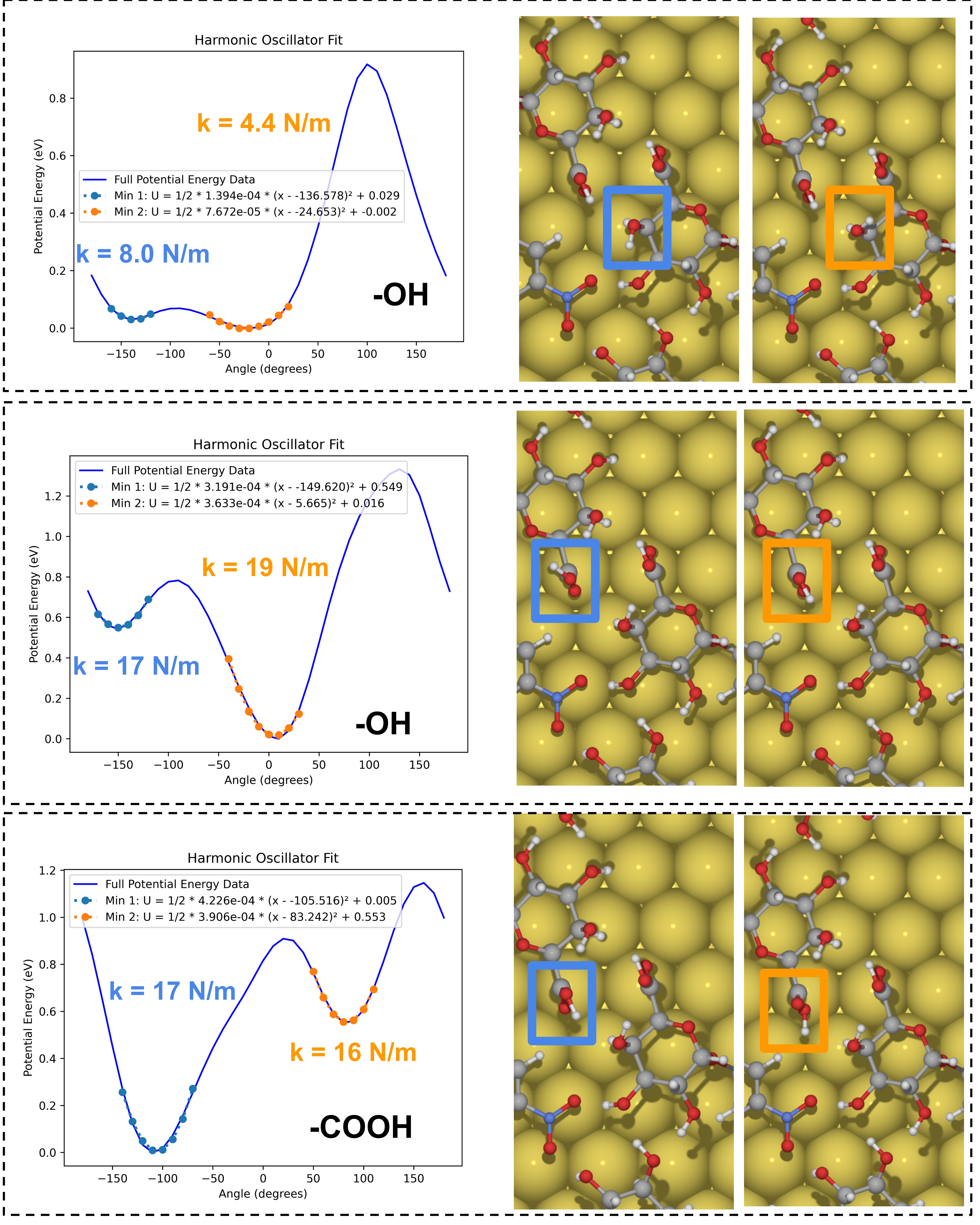}
    \caption{Determination of force constants from the -OH and -COOH group rotational motion bending potentials. The structures shown correspond to the minima in the potential energy plots as indicated by coloring. Note the different energy scales on the figures.}
    \label{fig:alpha_force_constants}
\end{figure}

To further investigate the effects of flexible groups of the sample on the image contrast of the NADG structure, we scanned the DFT energy for the rotation of the -OH groups in the central molecules of the monolayer as shown in Figure \ref{fig:alpha_OH_rotational_energy_analysis}. This was done due to the characteristic features appearing here at close tip-sample distances. 
Starting from the model structure corresponding the simulated images in the main manuscript, we scanned the DFT energies corresponding to rotation of the left (d4, in green) and bottom (d1, in orange) -OH groups, which are equivalent to the right and top groups by symmetry. Here we find that perturbing the bottom -OH group has two minima at d1 = \textminus150$^{\circ}$ and \textminus20$^{\circ}$, separated by a 0.1 eV barrier. This indicates that the -OH groups at the top and bottom of the central structure are quite flexible, having a range of 130 degrees of rotation within a small energy window. We also scanned the rotations of the -OH groups while keeping the other -OH group in the higher-lying local minima, but found that the initial rotation of d1 is the most facile alteration of the structure as determined by the barriers, even though a smaller rotational barrier is found for the transition of d1 from 10$^{\circ}$ to 90$^{\circ}$, displaying a computed barrier of 0.02 eV, this requires that d4 is first changed from 10$^{\circ}$ to \textminus150$^{\circ}$, the two latter minima being separated by a 0.8 eV barrier. However, this analysis assumes that single bonds are rotated in isolation, while in reality the whole system hydrogen-bonded network are affected by localized perturbations in the structure, meaning that this analysis is just a simple model trying to explain the contrast changes at close tip-sample distances.     
We also estimated the rotational forces by numerical derivation of the DFT potential energy U:
\[F = -dU/\theta\]
as well as the activation forces ($F_{activation}$) required to surmount the energy barriers for the transitions between local rotational minima using the linear distance ($x_{linear}$) the hydrogen has to traverse around the oxygen atom,
\[x_{linear} = (2\pi\cdot r\cdot\theta)/360\]
\[F_{activation} = \Delta E/x_{linear}\]

where the $\Delta E$ is the energy barrier height in J, r is the O--H bond distance in m, $\theta$ is the angle of rotation between the minimum and the barrier. These are found to vary between 0.05 -- 1.2 nN for the lowest and highest OH-rotational barriers, respectively. To facilitate comparison with the bending stiffness of the CO tip (typical values of k = 0.24 -- 0.5 N/m)\cite{hapala_mapping_2016}, we approximated the -OH and -COOH group rotational minima as harmonic oscillators to estimate their bending stiffness values as shown in Figure \ref{fig:alpha_force_constants}. This was done by fitting a second degree polynomial to the potential energy minima,
\[U = 1/2\cdot kx_{linear}^2\]
Here we find force constants for the rotationally bound minima ranging from k = 4.4 -- 19 N/m, which are about one and two orders of magnitude larger than the bending stiffness of the CO-tip. This suggests that the bending of the CO tip alone is not enough to perturb these functional groups significantly, but given close enough tip-sample distances, the C--O stretching component could play a significant role with a force constant around 1860 N/m for the isolated molecule. Furthermore, the rotation stiffness of the -COOH group is slightly smaller than the rotation of the left/right -OH groups, which are bound to it, which implies that tip-induced rotation of the latter is possibly accompanied by rotation of the former to some extent.  

\clearpage

\addcontentsline{toc}{subsection}{Orbital interactions}
\subsection{Orbital interactions} 

\begin{figure}[htbp]
    \centering
    \includegraphics[width=\textwidth]{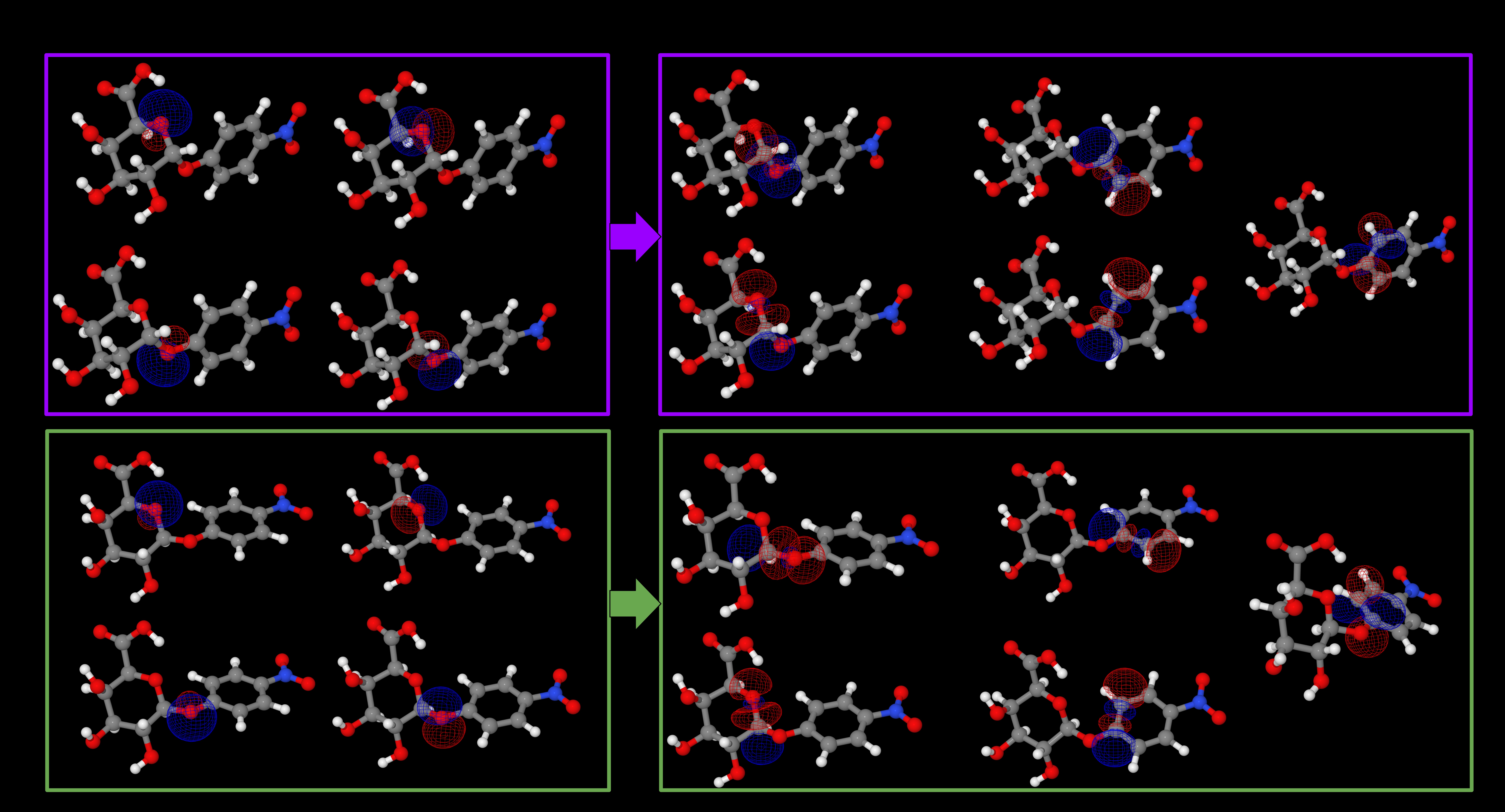}
    \caption{Overview of donor and acceptor orbitals from Natural Bond Orbital (NBO) analysis. The shown orbitals correspond to the pre-orthogonal Natural Bond Orbitals (PNBO), which retains the symmetry of the component atomic orbitals for intuitive visualization and analysis. Orbitals in purple boxes correspond to the $\alpha$-anomer, green to the $\beta$-anomer.}
    \label{fig:orbital_interactions}
\end{figure}

The donor-acceptor orbital interactions relevant to the anomeric effects for \textbf{NADG} and \textbf{NBDG} are displayed in Figure \ref{fig:orbital_interactions}, computed using the NBO method in Gaussian\cite{glendening_nbo_nodate,g16}. Typically, one considers the O$_5$/O$_1$ lone pair and the antibonding C$_1$--O$_1$/C$_1$--O$_5$ orbitals the most important for the anomeric effect. These correspond to NBOs 64, 65, 76, 77 for the O$_5$/O$_1$ $\sigma$- and $\pi$-lone pairs as donors, while the C$_1$--O$_1$/C$_1$--O$_5$ antibonding orbitals are described by NBOs 343 and 337 as acceptors, respectively. Furthermore, we find that interactions with $\sigma$*- and $\pi$*-orbitals in the nitrophenyl groups, NBOs 357, 358 and 359 are significant for the overall anomeric effect in the assembled structures, specifically by providing alternative acceptor sites for the O$_1$-lone pairs.    

\clearpage

\addcontentsline{toc}{subsection}{Charge density differences}
\subsection{Charge density differences} 

\newpage

\end{document}